\documentclass[aps,twocolumn,pra,reprint,amsmath,amssymb,floatfix,footinbib,superscriptaddress]{revtex4}

\usepackage{amsmath}	%allows the use of mathematical symbols
\usepackage{dsfont}  % 添加此包
\usepackage{gensymb}
\usepackage{hyperref}
\hypersetup{colorlinks=true}
\usepackage{upgreek}
\usepackage{graphics}
\usepackage{hyperref}
\usepackage{epsfig}
\usepackage{color}
\usepackage{bm}
\usepackage{float}
\usepackage{ulem} % \sout
\usepackage{blindtext}

\usepackage{CJK}        % For Windows Only
\usepackage{url}
\usepackage{multirow}
\newcommand{\ii}{\imath}   % imag unit

\begin{document}
% \begin{CJK*}{GBK}{kai}

\title{Stabilization of zigzag order in NiPS$_3$ via positive biquadratic interaction}

\author{Qiang Luo}
\email[]{qiangluo@nuaa.edu.cn}
\affiliation{College of Physics, Nanjing University of Aeronautics and Astronautics, Nanjing, 211106, China}
\affiliation{Key Laboratory of Aerospace Information Materials and Physics (NUAA), MIIT, Nanjing, 211106, China}

\author{Shuhang Yang}
\affiliation{College of Physics, Nanjing University of Aeronautics and Astronautics, Nanjing, 211106, China}

\author{Xiaoying Wang}
\affiliation{College of Physics, Nanjing University of Aeronautics and Astronautics, Nanjing, 211106, China}

\author{Mengdong Li}
\affiliation{College of Physics, Nanjing University of Aeronautics and Astronautics, Nanjing, 211106, China}

\author{Zhengyu Jiang}
\email[]{zyjiang2000@nuaa.edu.cn}
\affiliation{College of Physics, Nanjing University of Aeronautics and Astronautics, Nanjing, 211106, China}

\author{Chunlan Ma}
\affiliation{School of Physical Science and Technology, Suzhou University of Science and Technology, Suzhou 215009, China}

\author{Yan Zhu}
\email[]{yzhu@nuaa.edu.cn}
\affiliation{College of Physics, Nanjing University of Aeronautics and Astronautics, Nanjing, 211106, China}
\affiliation{Key Laboratory of Aerospace Information Materials and Physics (NUAA), MIIT, Nanjing, 211106, China}

\date{\today}

\begin{abstract}
  Despite extensive research, the precise spin Hamiltonian of the van der Waals antiferromagnet NiPS$_3$---which hosts a zigzag-ordered ground state---remains debated. While consensus has emerged on ferromagnetic nearest-neighbor ($J_1$) and antiferromagnetic third-nearest-neighbor ($J_3$) Heisenberg interactions, recent studies suggest that a biquadratic ($B$) exchange term may also play a role, though its estimated magnitude varies widely. To address this controversy, we perform density functional theory calculations and extract a positive biquadratic interaction with $B/J_3 \approx 0.44$. Within the minimal $J_1$-$J_3$-$B$ model, we show that these parameters naturally stabilize zigzag ordering using minimally augmented spin-wave theory. Density-matrix renormalization group calculations further validate our extracted parameters as a reasonable description of the ground state. Although fully resolving the spin Hamiltonian of NiPS$_3$ requires further investigation, our findings provide new insights into its biquadratic interaction.
\end{abstract}

\pacs{}
%\keywords{}

\maketitle

%%%%%%%%%%%%%%%%%%%%%%%%%%%%%%%%%%%%%%%%%%%%%%%%%%%%%%%%%%%%%%%%%%%%%%%%%%%%%%
\section{Introduction}

Honeycomb-lattice van der Waals antiferromagnets have emerged as a fertile ground for a remarkable array of magnetic, electronic, and optical quantum phenomena \cite{Burch2018Nat,Gong2017Nat,Huang2017Nat,Banerjee2017Sci,Zong2023Nat,Ilyas2024Nat}, due to the distinct lattice geometry with low coordination number and the substantial competing exchange interactions that engender intricate magnetic ground states. The zigzag-type antiferromagnetic (AFM) order, which represents the most commonly realized magnetic configuration, has been observed across a wide spectrum of crystal symmetries and transition metal species (for a systemic review, see Ref.~\cite{Garlea2025AC}). These materials accommodate magnetic ions ranging from $4d/5d$ elements with strong spin-orbit coupling (e.g., $\mathrm{Ir}^{4+}$ \cite{Singh2012PRL,Chun2015NP}, $\mathrm{Ru}^{3+}$ \cite{Plumb2014PRB,Banerjee2017Sci}) to $3d$ transition metals (e.g., $\mathrm{Co}^{2+}$ \cite{Yao2020PRB,Lin2021NC,Li2022PRX}, $\mathrm{Ni}^{2+}$ \cite{Gao2021PRB,Shangguan2023NP}), leading to effective spins that vary from $S_{\mathrm{eff}} = 1/2$ in Kramers doublet systems to integer spins in nickelates. This broad family of materials, united by their zigzag ground state but spanning diverse energy scales and anisotropy preferences, provides a unique opportunity to disentangle the microscopic origins of magnetic order and probe the proximity to Kitaev physics.

In particular, transition metal thiophosphates $M$PS$_3$ ($M$ = Ni, Co, Fe, Mn) stand out as a rich family of layered magnetic compounds \cite{Joy1992PRB,Sivadas2015PRB,Chittari2016PRB,Coak2019JPCM}, within which NiPS$_3$ that hosts a zigzag ordering has attracted particular interest as it has the smallest spin ($S = 1$) while exhibiting the highest N{\'e}el temperature ($T_N \approx 155$ K \cite{Kim2019NC,Plumley2024npjQM}) among its isostructural counterparts. The ordered magnetic moments lie predominantly in-plane along an easy-axis \cite{Lancon2018PRB,Wildes2022PRB,Scheie2023PRB}, with a small tilt of $8 \sim 15^{\circ}$ from the basal plane \cite{Wildes2015PRB,Lane2020PRB,Lee2024AFM}. Many intriguing quantum phenomena have been observed in NiPS$_3$, including coherent many-body exciton \cite{Kang2020Nat}, anomalous scattering signatures \cite{Scheie2023PRB}, field-induced spin-flop transition \cite{Basnet2021PRM,Bougamha2022PRB},
and enhanced thermal conductivity due to spin-lattice coupling \cite{Liu2025ACSANM}. A precise understanding of these phenomena, however, depends critically on an accurate description of the underlying spin Hamiltonian.

To date, two prevailing theoretical mechanisms have been proposed to stabilize zigzag order in honeycomb lattices: the conventional $J_1$-$J_2$-$J_3$ Heisenberg model with ferromagnetic (FM) $J_1$ and AFM $J_3$ \cite{Rastelli1979PBC,Bose2023PRB,Jiang2023PRB,Chernyshev2025PRB}, and the extended Kitaev model incorporating bond-dependent Kitaev interaction and off-diagonal $\Gamma$, $\Gamma'$ terms \cite{Rau2014PRL}. These competing frameworks are exemplified by the field-induced one-third magnetization plateau observed above the zero-field zigzag order in Na$_3$Ni$_2$BiO$_6$ \cite{Shangguan2023NP}. While the initial proposal attributed this plateau to Kitaev frustration \cite{Shangguan2023NP}, a recent density functional theory (DFT) calculation suggests that it originates from Heisenberg couplings $J_1$ and $J_3$ reinforced by out-of-plane single-ion anisotropy \cite{Konieczna2026arXiv}.
The difficulty in unambiguously identifying the spin Hamiltonian is also epitomized by NiPS$_3$, a subject of considerable debate for decades. A consensus has now emerged that a Heisenberg model extending to third nearest neighbors, augmented by single-ion anisotropy, captures its essential magnetism \cite{Lancon2018PRB,Wildes2022PRB,Wang2025JMMM}. However, the validity of this minimal model has been called into question by recent studies employing a two-band half-filled Hubbard model, which predict the presence of a biquadratic interaction \cite{Mellado2023APL}. Although two independent DFT calculations confirm its existence, they produce inconsistent estimates of its strength relative to the dominant $J_3$, with the ratio ranging from $-0.10$ \cite{Li2024PRB} to $-0.66$ \cite{Amirabbasi2024PRM}. This striking discrepancy underscores the need for further scrutiny to resolve the true nature of the magnetic interactions in NiPS$_3$.

\begin{figure}[!ht]
\centering
\includegraphics[width=0.95\columnwidth, clip]{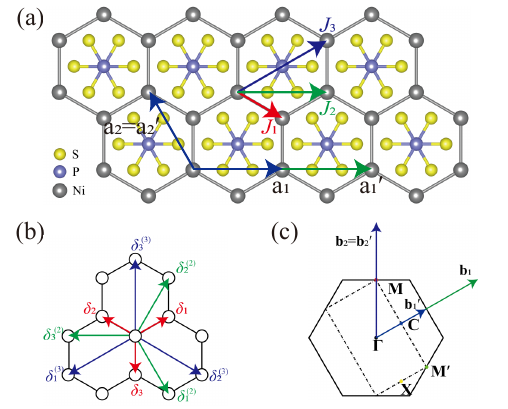}\\
\caption{(a) Crystal structure of NiPS$_3$. The red, green, and blue arrows denote the Heisenberg interactions of the first three neighbors, respectively. (b) Sketch of the honeycomb lattice with the first three nearest-neighbor vectors, $\boldsymbol{\delta}_{\gamma}$ (red), $\boldsymbol{\delta}_{\gamma}^{(2)}$ (green), and $\boldsymbol{\delta}_{\gamma}^{(3)}$ (blue), with $\gamma = x(1)$, $y(2)$, and $z(3)$.  (c) Brillouin zones of the magnetic primitive cell and the supercell. The dashed outline indicates the Brillouin zone of the zigzag supercell. The high-symmetry points along $\boldsymbol{\Gamma}$-$\textbf{X}$-$\textbf{M}'$-$\textbf{C}$-$\textbf{M}$ define the $K$-path for the band structure. $\textbf{b}_1$ and $\textbf{b}_2$, and $\textbf{b}_1'$ and $\textbf{b}_2'$ are the reciprocal lattice vectors of the primitive cell and the supercell, respectively.}\label{FIG-NiPS3}
\end{figure}

In this work, we aim to elucidate the role of the biquadratic interaction within the minimal $J_1$-$J_3$-$B$ model for NiPS$_3$ through a multi-faceted approach. We first perform detailed DFT calculations, extracting the exchange couplings through different combinations of linear and nonlinear spin configurations. Our analysis yields a positive biquadratic interaction, with a ratio of $B/J_3 \approx 0.44$. To validate this result, we employ minimally-augmented spin-wave theory (MASWT) \cite{Chernyshev2025PRB,Wenzel2012PRL,Coletta2013PRB,Coletta2014PRB} and demonstrate that the zigzag phase possesses a lower ground-state energy than the competing AFM phase. This finding is further corroborated by density matrix renormalization group (DMRG) calculations \cite{White1992PRL,Peschel1999,Schollwock2005RMP} performed on a 24-site hexagonal cluster, providing strong support for our proposed model.

%%%%%%%%%%%%%%%%%%%%%%%%%%%%%%%%%%%%%%%%%%%%%%%%%%%
\begin{table*}
\centering
\caption{Summary of representative exchange parameters and single-ion anisotropy for monolayer NiPS$_3$ (in meV). $\Delta_a$ represents the XXZ-type anisotropy. The data sets I–III include the biquadratic interaction $B$.}\label{TAB:NiPS3}
\begin{ruledtabular}
\begin{tabular}{ccccccccc}
  Set & Reference & $J_1$         & $J_2$ &$J_3$   & $\Delta_a$ & $B$ & $A_x$ & $A_z$ \\
  \hline %\\
  - & Lancon \textit{et al.} (2018) \cite{Lancon2018PRB}  & $-3.8$ & $0.2$ & $13.8$ & $1$ &$0$  &$0$ &$-0.3$ \\
  - & Kim \textit{et al.} (2019) \cite{Kim2019NC} & $3.18$ & $4.82$ & $9.08$ & $0.66$ & $0$ & $-0.89$ & $2.85$ \\
  - & Wildes \textit{et al.} (2022) \cite{Wildes2022PRB} & $-2.6(2)$ & $0.2(1)$ & $13.5(3)$ & $1$ & $0$ & $-0.01(1)$ & $0.21(8)$ \\
  - & Wang and Cai (2025) \cite{Wang2025JMMM} & $-4.16$ & $-0.27$ & $21.52$ & $1$ & $0$ & $0$ & $0$ \\
  \hline
  I & Li \textit{et al.} (2024) \cite{Li2024PRB} & $-3.38$ & $-0.64$ & $15.79$ & $1$ & $-1.66$ & $0$ & $1.25$ \\
  II & Amirabbasi and Kratzer (2024) \cite{Amirabbasi2024PRM} & $-2.46$ & $-0.14$ & $11.58$ & $1$ & $-6.91$ & $-0.22$ & $0$ \\
  III & \textbf{This work} & $-3.13$ & $-0.32$ & $14.58$ & $1$ & $6.49$ & $0$ & $0.16$
\end{tabular}
\end{ruledtabular}
\end{table*}

%%%%%%%%%%%%%%%%%%%%%%%%%%%%%%%%%%%%%%%%%%%%%%%%%%%%%%%%%%%%%%%%%%%%%%%%%%%%%%
\section{DFT Calculation: Methods and Model}

\subsection{Computational Methods}

The first-principles DFT calculations for geometric optimization and electronic structure of monolayer NiPS$_3$ were performed using the projector augmented-wave (PAW) method~\cite{Blochl1994PRB}, as implemented in the DS-PAW package and the Vienna ab initio simulation package (VASP)~\cite{Kresse1996PRB,Kresse1999PRB}.
The exchange-correlation functional was treated within the generalized gradient approximation (GGA) of the Perdew-Burke-Ernzerhof type~\cite{Perdew1996PRL}. To accurately capture the strong electron correlations in the localized Ni-3$d$ orbitals, the GGA+$U$ method was applied with an effective Hubbard parameter of $U_{\text{eff}} = 4.0$ eV~\cite{Gu2019PRB,Wang2025JMMM,Li2024PRB}, a value widely adopted and validated in previous first-principles studies of monolayer NiPS$_3$. The wave functions were expanded using a plane-wave basis set with a kinetic energy cutoff of 380 eV. The first Brillouin zone was sampled using a $13\times13\times1$ Monkhorst-Pack $k$-point mesh for the unit cell, while a $7\times13\times1$ mesh was adopted for the $2\times1\times1$ supercell. All atomic positions were fully relaxed towards the ground state until the residual forces on each atom were less than $10^{-2}$ eV/\AA. The convergence criterion for the total energy was strictly set to $10^{-6}$ eV, and the final total energy for the full supercell converged to $-8407.78$ eV in the DS-PAW calculation. A lattice constant of $a = 5.86$ \AA~was adopted~\cite{Li2024PRB}, and a Gaussian smearing width of 0.05 eV was applied to accelerate convergence during the structural relaxation and static calculations. Furthermore, the monolayer NiPS$_3$ was modeled using an atomic slab geometry with periodic boundary conditions, incorporating a vacuum region of 20 \AA~between adjacent layers to eliminate effects from interlayer interactions.

\begin{figure}[!ht]
\centering
\includegraphics[width=0.95\columnwidth, clip]{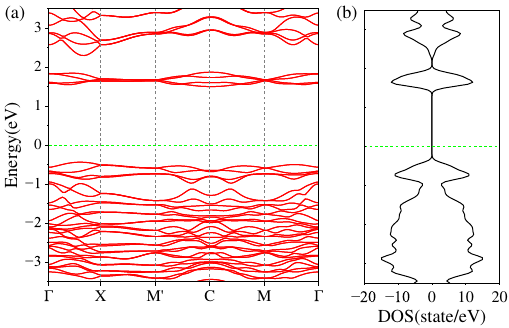}\\
\caption{(a) Electronic structure and (b) DOS of monolayer NiPS$_3$ calculated by DS-PAW software. The high-symmetry points in the reciprocal space are indicated in Fig.~\ref{FIG-NiPS3}(c). The horizontal dotted lines at zero energy are guides to the eye.}\label{FIG-DSEgvsDOS}
\end{figure}

%%============================================================================
\subsection{Electronic structure and Spin Hamiltonian}

The crystal structure of NiPS$_3$ is shown in Fig.~\ref{FIG-NiPS3}(a), where two Ni atoms (effective spin $S=1$) occupy two interpenetrating triangular sublattices. Each Ni atom is located at the center of a regular octahedron, collectively forming a honeycomb-like hexagonal lattice. The $K$-path adopted for the electronic band structure calculations traces along the high-symmetry points within the irreducible Brillouin zone, as depicted in Fig.~\ref{FIG-NiPS3}(c).

Figure~\ref{FIG-DSEgvsDOS} presents the electronic structure and density of states (DOS) of NiPS$_3$. The valence band maximum is located along the $\boldsymbol{\Gamma}$-$\textbf{X}$ path, whereas the conduction band minimum resides at the $\textbf{C}$ point, characterizing NiPS$_3$ as an indirect-band-gap semiconductor with a gap of 1.94 eV. This value is consistent with the previous numerical calculation and experimental measurements \cite{Du2016ACSNano,Kim2018PRL,Lane2020PRB,Xu2025OE}. Notably, the relatively flat band dispersion near the Fermi level, manifested as sharp peaks in the DOS, indicates highly localized electronic states. Such features suggest that the electronic structure is highly susceptible to external perturbations, offering a promising platform for band engineering via strain, electric, or magnetic fields.

The microscopic Hamiltonian of NiPS$_3$ has been widely studied over the last decade. Hitherto, the most generic model for NiPS$_3$ is proposed as
\begin{align}\label{EQ:HamNiPS3}
\mathcal{H} =& \sum_{\langle i,j\rangle_1}[ J_{1} \textbf{S}_{i}\cdot \textbf{S}_{j} + B(\textbf{S}_{i}\cdot \textbf{S}_{j})^{2}] + \sum_{\langle i,k\rangle_2} J_{2} \textbf{S}_{i} \cdot \textbf{S}_{k} \nonumber \\
 & +\sum_{\langle i,l\rangle_3} J_{3} \textbf{S}_{i} \cdot \textbf{S}_{l} + \sum_{i} \left[A_{x}(S_{i}^x)^{2} + A_{z}(S_{i}^z)^{2} \right]
\end{align}
in which $\textbf{S}_i = (S_i^x, S_i^y, S_i^z)$ is the spin-1 operator. $J_1$, $J_2$, and $J_3$ represent the first, second, and third nearest-neighbor Heisenberg exchange parameters, $B$ denotes the strength of biquadratic interaction, and $A_x$ and $A_z$ are the single-ion anisotropy terms. 
To extract the Heisenberg and biquadratic exchange interactions using the four-state energy mapping method with spin-orbit coupling, we consider both collinear and non-collinear magnetic configurations. Specifically, we examine the FM, N\'eel-type AFM, stripe-AFM, and zigzag-AFM states, as well as three non-collinear configurations with relative spin angles of 30$^\circ$, 90$^\circ$ and 120$^\circ$, as illustrated in Sec. S1 in the supplemental material \cite{SuppMat}. 

The Heisenberg exchange parameters can be obtained by fitting the energies of three linear magnetic configurations relative to the FM reference, yielding $(J_1, J_2, J_3) \approx (-3.2091, -0.2791, 14.5556)$~meV. Among these, $J_3$ is found to be the dominant interaction. By further including three nonlinear magnetic configurations in the fitting, we obtain an improved set of parameters, $(J_1, J_2, J_3, B) \approx (-3.1309, -0.3182, 14.5817, 6.4868)$ meV. Notably, the fitted Heisenberg exchanges remain rather stable, and the biquadratic interaction $B$ is found to be positive. Nevertheless, alternative combinations of linear and nonlinear magnetic configurations can lead to different results. 
However, caution is needed because four of the six configurations happen to share identical coefficients for $J_1$ and $J_3$ (see Sec. S1 in the supplemental material \cite{SuppMat}). To avoid underfitting, the remaining two stripe-AFM and zigzag-AFM configurations---which have distinct prefactors---must be included to uniquely resolve all exchange parameters.

\begin{figure}[!ht]
\centering
\includegraphics[width=0.95\columnwidth, clip]{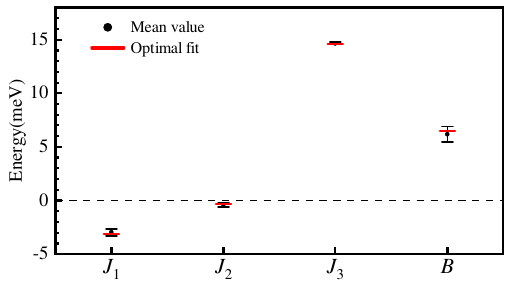}\\
\caption{Estimates of the exchange parameters $J_1$, $J_2$, $J_3$, and $B$ obtained from different sets of magnetic configurations. The optimal fits (red) are derived using all six linear and nonlinear configurations, while the mean values and error bars (black) are obtained from a statistical analysis of selected subsets.}\label{FIG-FittedCoupling}
\end{figure}

Figure~\ref{FIG-FittedCoupling} displays the estimated mean values and standard errors of $\{J_1, J_2, J_3, B\}$ obtained from this reduced set. Our best estimate lies within the resulting error bars, confirming the reliability of the extracted exchange parameters. In addition, the single-ion anisotropy $A_z$ is evaluated from the magnetocrystalline anisotropy energy (MAE), defined as the energy difference $E_{100} - E_{001}$, giving a tiny value of $A_z = 0.16$~meV.

To assess the robustness of our results against the choice of Hubbard $U_{\mathrm{eff}}$, we performed additional DFT calculations with $U_{\mathrm{eff}} = 3.0$~eV and $5.0$~eV and compared them to the original $U_{\mathrm{eff}} = 4.0$~eV results. While all extracted exchange parameters decrease monotonically in magnitude with increasing $U_{\mathrm{eff}}$, their ratios remain confined to a narrow range, as can be seen from Sec. S1 in the supplemental material \cite{SuppMat}. Notably, $B$ stays positive for all three $U_{\mathrm{eff}}$ values, and the ratio $B/J_3$ is stable at $0.44^{+0.03}_{-0.02}$. 
We have further verified the robustness of our findings by performing additional calculations without spin-orbit coupling and using the HSE06 hybrid functional \cite{Franchini2007PRB}. In both cases, the extracted magnetic parameters, particularly the sign and magnitude of $B$, remain largely unchanged. 
These findings demonstrate that both the extracted parameters and our main physical conclusions are robust against variations in $U_{\mathrm{eff}}$, spin-orbit coupling, and the choice of exchange-correlation functional.

The fitted parameters for NiPS$_3$ are also summarized in Table~\ref{TAB:NiPS3}, along with representative data sets from the literature \cite{Lancon2018PRB, Kim2019NC, Wildes2022PRB, Wang2025JMMM, Li2024PRB, Amirabbasi2024PRM}.
It is observed from these data sets that $J_3$ is the dominant magnetic interaction. Physically, the strong $J_3$ can be consistently understood from both direct-exchange and superexchange perspectives. The direct-exchange mechanism, arising from the fully occupied $t_{2g}$ subshell, yields dominant hopping via odd orbitals at third nearest-neighbor distances \cite{Autieri2022JPCC}. Concurrently, a super-superexchange path via sulfur $p$ orbitals, particularly involving Ni $e_g$ states, also produces a large $J_3$ \cite{Leon20252DMater}. Thus, both approaches robustly explain the unusually large magnitude of $J_3$ in this material.
For data sets I, II, and III, which include the biquadratic interaction, it is striking that they consistently yield $J_1/J_3 \approx -0.21$ and a negligibly small $J_2/J_3$. However, the reported biquadratic interactions show considerable scatter, both in magnitude and sign. This indicates that further many-body calculations are warranted to distinguish among these different proposals.

%%%%%%%%%%%%%%%%%%%%%%%%%%%%%%%%%%%%%%%%%%%%%%%%%%%%%%%%%%%%%%%%%%%%%%%%%%%%%%
\section{Results and Discussion}

To establish a comprehensive understanding of the microscopic spin Hamiltonian, we performed a hierarchical study combining classical Monte Carlo simulations, spin-wave analysis, and DMRG calculations.

\begin{figure}[!ht]
\centering
\includegraphics[width=0.95\columnwidth, clip]{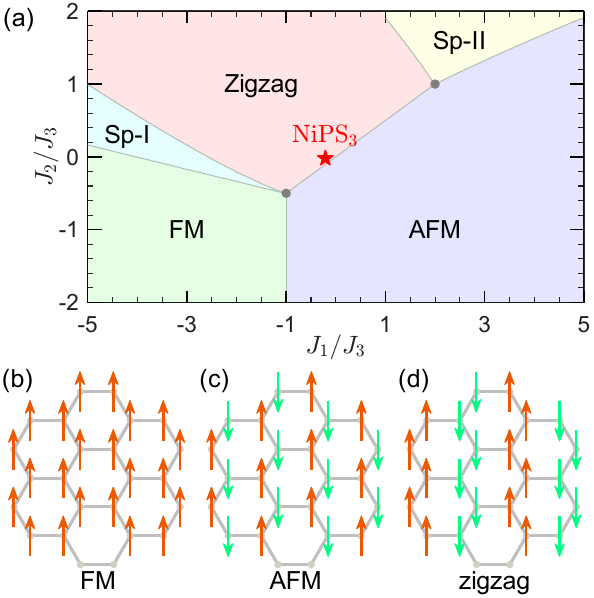}\\
\caption{(a) Classical phase diagram of the $J_1$-$J_2$-$J_3$ model in the ($J_1/J_3$, $J_2/J_3$) parameter space with $J_3 = 1$. This phase diagram was mapped out by the energy comparison of three collinear phases (FM, AFM, and zigzag) and two proposed noncollinear spiral phases. Parameters specific to NiPS$_3$ are highlighted with a pentagram ($\star$). (b) Spin configurations of the FM, AFM, and zigzag phases, from the left to right.}\label{FIG-J1J2J3CPD}
\end{figure}

%%============================================================================
\subsection{Classical energy minimization}

We begin by sketching the classical phase diagram of the $J_1$-$J_2$-$J_3$ model \cite{Rastelli1979PBC,Fouet2001EPJB}, taking $J_3 = 1$ as the unit of energy. In the pure $J_3$ limit ($J_1 = J_2 = 0$), the system decouples into independent AFM sublattices. While each sublattice---a bipartite honeycomb lattice---exhibits N\'eel order, the absence of inter-sublattice coupling leaves the relative orientation of their order parameters entirely free, resulting in a continuous ground-state manifold. Upon introducing small $J_1$ and/or $J_2$ terms, the system selects either the AFM or the zigzag phase. For negative and moderate $J_1$ and $J_2$ interactions, the FM phase becomes the ground state. The corresponding classical energies per site are given by
\begin{subequations}\label{EQ:ClEg}
\begin{equation}
E_{cl}^{FM} = S^2\left[\frac{3}{2}(J_1 + 2J_2 + J_3) + \frac{3}{2}BS^2\right],
\end{equation}
\begin{equation}
E_{cl}^{AFM} = S^2\left[\frac{3}{2}(-J_1 + 2J_2 - J_3) + \frac{3}{2}BS^2\right],
\end{equation}
\begin{equation}
E_{cl}^{ZZ} = S^2\left[\frac{1}{2}(J_1 - 2J_2 - 3J_3) + \frac{3}{2}BS^2\right].
\end{equation}
\end{subequations}
In Eq.~\eqref{EQ:ClEg}, we include the energy contribution from the biquadratic interaction for completeness. Owing to its isotropic nature, this term contributes equally to all phases, implying that it does not drive additional phase transitions among them.

Beyond these simple collinear phases, the ground-state phase diagram hosts several noncollinear phases characterized by a variable ordering wavevector. Following the proposal by Rastelli \textit{et al.} \cite{Rastelli1979PBC}, single-$Q$ spiral phases provide a reasonable approximation. For the spiral-I phase, the ordering wavevector is $\mathbf{Q} = (0, Q_y)$, with
\begin{equation}\label{EQ:OWV_Qy}
Q_y^{\pm} = \frac{2}{3} \textrm{acos} \left(\frac{1}{2J_3(J_1+2J_3)} \big[(J_1+3J_3) \pm\sqrt{F}\big]\right)
\end{equation}
where 
$F = \frac{(2J_2)^2 (J_1+J_3)^3 J_3(J_1-2J_3)}{(2J_2)^2 - J_3(J_1+2J_3)}$.
The spins on the two sublattices are canted relative to each other by an angle
$\phi^{\pm} = -\frac{1}{2} Q_{y}^{\pm} + \textrm{atan} \left(\frac{ (J_1+J_3)\sin(3Q_y/2)}{2J_1 + (J_1+3J_3)\cos(3Q_y/2)}\right)$. The classical energy per site is $E_{\rm{Sp-I}} = \min(E_{\rm{Sp-I}}^{+}, E_{\rm{Sp-I}}^{-})$, in which
\begin{align}\label{EQ:ClEgSpI}
E_{\rm{Sp-I}}^{\pm} =& \frac{1}{2} \Big[J_1\big(\cos(Q_y^{\pm}-\phi^{\pm}) + 2\cos(Q_y^{\pm}+\phi^{\pm})\big) \nonumber\\
&+ 2J_2\big(1+2\cos(3Q_y^{\pm}/2)\big) \nonumber\\
&+ J_3\big(\cos(2Q_y^{\pm}+\phi^{\pm}) + 2\cos(Q_y^{\pm}-\phi^{\pm})\big)\Big].
\end{align}

For the spiral-II phase, the ordering wavevector is $\mathbf{Q} = (Q_x, 0)$, with
\begin{equation}\label{EQ:OWV_Qx}
Q_x = \frac{2}{\sqrt{3}} \textrm{acos} \left(-\frac{J_1+2J_2}{4(J_2+J_3)}\right),
\end{equation}
and the corresponding classical energy per site is
\begin{equation}
E_{\rm{Sp-II}} = -\frac{3}{2} \left[J_2 + \frac{(J_1-2J_3)^2}{12(J_2+J_3)}\right].
\end{equation}
The physical requirement that the argument of the arccosine in Eqs.~\eqref{EQ:OWV_Qy} and \eqref{EQ:OWV_Qx} be no greater than unity in absolute value constrains the possible values of $\mathbf{Q}$, thereby delineating the phase boundaries of these spiral phases. Comparing the classical energies of the five phases yields the classical phase diagram shown in Fig.~\ref{FIG-J1J2J3CPD}(a), with representative spin configurations for the FM, AFM, and zigzag phases illustrated in Figs.~\ref{FIG-J1J2J3CPD}(b)-\ref{FIG-J1J2J3CPD}(d).

\begin{figure}[!ht]
\centering
\includegraphics[width=0.95\columnwidth, clip]{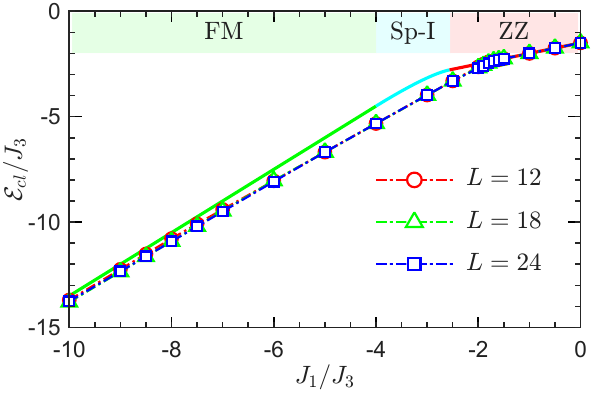}\\
\caption{Classical energy per site $\mathcal{E}_{cl} = E_{cl}/N$ as a function of $J_1/J_3$ in the $J_1$-$J_3$ model, obtained from Monte Carlo simulations on $2\times L \times L$ clusters with $L = 12$ (red circles), 18 (green triangles), and 24 (blue squares). Solid lines represent the exact energy expressions for the FM, spiral-I, and zigzag phases. Inset: Classical phase diagram extracted from Fig.~\ref{FIG-J1J2J3CPD}.
}\label{FIG-CMC}
\end{figure}

However, the proposed spiral phases do not constitute the true ground states, as they are energetically higher than incommensurate or large-unit-cell phases identified via classical Monte Carlo simulations \cite{Metropolis1953JCP,Hukushima1996JPSJ}. To demonstrate this, we focus on the $J_1$-$J_3$ limit (with $J_2 = 0$). Comparing the energies of the FM and zigzag phases [Eq.~\eqref{EQ:ClEg}] with that of the spiral-I phase [Eq.~\eqref{EQ:ClEgSpI}], we find that the spiral-I phase persists in the interval $-4 < J_1/J_3 < 8/(\sqrt{17} - 1)$, while the zigzag phase emerges in the range $8/(\sqrt{17} - 1) < J_1/J_3 < 0$ (see Fig.~\ref{FIG-CMC}).
We further performed large-scale parallel tempering Monte Carlo simulations for system sizes up to $N = 1152$, accumulating $10^6$ Monte Carlo samples as a function of $J_1/J_3 \in [-10, 0]$. As shown in Fig.~\ref{FIG-CMC}, upon increasing the FM coupling $J_1$ from zero, the region of the zigzag phase shrinks to $-1.65 \lesssim J_1/J_3 < 0$.
Beyond this region, an incommensurate phase emerges and persists at least down to $J_1/J_3 = -10$, despite its energy approaching that of the FM phase. Our results suggest that the classical phase diagram is considerably more intricate than previously anticipated. Nevertheless, they confirm the persistence of the zigzag phase in the parameter regime relevant to NiPS$_3$.

%%============================================================================
\subsection{Spin-wave energy}

Given that NiPS$_3$ resides in close proximity to the phase boundary separating the AFM and zigzag phases [see Fig.~\ref{FIG-J1J2J3CPD}(a)], we now establish a unified framework for deriving the spin-wave spectra of both magnetic orders.
Without loss of generality, we assume the classical ordered moment to be oriented along the unit vector $\mathbf{n}$. Accordingly, we define $\tilde{S}_i^z = \mathbf{S}_i \cdot \mathbf{n}$, and introduce the ladder operators $\tilde{S}_i^\pm \equiv (\mathbf{S}_i \cdot \mathbf{e}) \pm i [\mathbf{S}_i \cdot (\mathbf{n} \times \mathbf{e})]$, where $\mathbf{e}$ is an arbitrary unit vector perpendicular to $\mathbf{n}$ satisfying the right-hand rule \cite{Janssen2019JPCM}.
Using the Holstein-Primakoff transformation, the spin operators are expressed in terms of bosonic creation and annihilation operators $a_i^\dagger$ and $a_i$ as
$
\tilde{S}_i^+ \simeq \sqrt{2S} \, a_i,
\tilde{S}_i^- \simeq \sqrt{2S} \, a_i^\dagger,
\tilde{S}_i^z = S - a_i^\dagger a_i
$.
The spin operator at site $i$ can then be written as
\begin{eqnarray}\label{SpinOpN_E_NxE}
\mathbf{S}_{\tau,i}
&=& \sqrt{\frac{S}{2}}(a_i+a_i^{\dagger}){\mathbf{e}} + \tau\sqrt{\frac{S}{2}}(-\ii a_i+\ii a_i^{\dagger}) (\mathbf n \times \mathbf e) \nonumber\\
& & +\tau(S-a_i^{\dagger}a_i) {\mathbf{n}}
\end{eqnarray}
where $\tau = +1$ ($-1$) denotes spins aligned (antialigned) with $\mathbf{n}$ in the classical ground state \cite{Luo2021NPJ}.
The $\gamma$-component of the spin is given by $S_{\tau,i}^{\gamma} = \mathbf{S}_{\tau,i} \cdot \mathbf{e}_{\gamma}$, with $\mathbf{e}_{\gamma}$ being the cubic spin-space basis vectors. Given that only $A_x$ and $A_z$ single-ion anisotropies are present in Eq.~\eqref{EQ:HamNiPS3}, the ordered moments are confined to the $xz$ plane. In this geometry, the spin components in the laboratory frame are related to those in the local frame via
$
(S_i^x, S_i^y, S_i^z) = (\sin\theta \, \tilde{S}_i^z + \cos\theta \, \tilde{S}_i^x,\; \tilde{S}_i^y,\; \cos\theta \, \tilde{S}_i^z - \sin\theta \, \tilde{S}_i^x)
$.

By substituting the Holstein-Primakoff transformation into the Hamiltonian in Eq.~\eqref{EQ:HamNiPS3} and subsequently performing a Fourier transform, we arrive at the following linear spin-wave Hamiltonian \cite{Chernyshev2025PRB}:
\begin{eqnarray}\label{HswZZ}
\mathcal{H}_{\mathrm{SW}} = E_{cl} + \frac{S}{2} \sum_{\bm q}\bigg(
    \hat{\bf x}_{\mathbf{q}}^\dagger
    \hat{\bf H}_{\mathbf{q}}
    \hat{\bf x}_{\mathbf{q}}^{\phantom{\dagger}}
    - \frac{1}{2} \textrm{tr}\left( \hat{\bf H}_{\mathbf{q}} \right) \bigg)
\end{eqnarray}
where $E_{cl}$ is the classical energy, $\hat{\bf x}^\dag_{\mathbf{q}} = \big( a^\dag_{\mathbf{q}}, b^\dag_{\mathbf{q}}, \cdots,
a^{\phantom \dag}_{\mathbf{-q}},b^{\phantom \dag}_{\mathbf{-q}}, \cdots\big)$
is a Nambu spinor of dimension $2 n_s$, and $\hat{\bf H}_{\mathbf{q}}$ denotes the $2 n_s\times 2 n_s$ Bogoliubov–de Gennes (BdG) Hamiltonian taking the form
\begin{eqnarray}
\hat{\bf H}_{\mathbf{q}} =
\left(\begin{array}{cc}
	\hat{\Lambda}_{\mathbf{q}}              &    \hat{\Delta}_{\mathbf{q}}       \\
    \hat{\Delta}_{\mathbf{q}}^{\dagger}     &    \hat{\Lambda}_{-\mathbf{q}}^T
\end{array}\right).
\end{eqnarray}
Here, $n_s$ denotes the number of sublattice within each magnetic unit cell. For brevity, we present only the main results and relegate the detailed general derivation of the BdG formalism for collinear states in the two- and four-sublattice cases to Sec. S2 in the Supplemental Material \cite{SuppMat}.

For the two-sublattice AFM phase, $E_{cl} =
N S^2 (-3J_{1} + 6J_{2} - 3J_{3})/2 + 3N S^4 B/2 + N [A_z \left(\frac{S}{2} \sin^2\theta + S^2 \cos^2\theta\right) + A_x (\frac{S}{2} \cos^2\theta + S^2 \sin^2\theta)]$. $\theta$ is the polar angle of the magnetic moment, obtained by minimizing the classical energy. The matrices $\hat{\Lambda}_{\mathbf{q}}$ and $\hat{\Delta}_{\mathbf{q}}$ take the block-diagonal forms

\begin{eqnarray}\label{HtotAACC}
\hat{\Lambda}_{\bf q} =
\left(\begin{array}{cc}
	\mathbb{A}_{\mathbf q}                  & 0      \\
    0  & \mathbb{A}_{\mathbf q}             \\
\end{array}\right),\quad
\hat{\Delta}_{\bf q} =
\left(\begin{array}{cc}
	\mathbb{C}_{\mathbf q}       & \mathbb{D}_{\mathbf q}     \\
    \mathbb{D}_{-\mathbf q}      & \mathbb{D}_{\mathbf q}    \\
\end{array}\right),
\end{eqnarray}
where the matrix elements are given by
$\mathbb{A}_{\mathbf q} = A_z (1 - 3\cos^2\theta) + A_x (1 - 3\sin^2\theta) - \left(-3J_{1} + 6J_{2} - 3J_{3} + 6S^2 B\right) + J_{2} \sum_{\gamma=x,y,z} (e^{i{\mathbf q} \boldsymbol\delta^{(2)}_{\gamma}}+e^{-i{\mathbf q} \boldsymbol\delta^{(2)}_{\gamma}})$,
$\mathbb{C}_{\mathbf q} = \frac{1}{2} (A_z \sin^2\theta + A_x \cos^2\theta)$,
and $\mathbb{D}_{\mathbf q} = J^{-}_{1}\sum_{\gamma=x,y,z} e^{i{\mathbf q}\boldsymbol\delta_{\gamma}} + J_{3}\sum_{\gamma=x,y,z} e^{i{\mathbf q} \boldsymbol\delta^{(3)}_{\gamma}}$.
Here, $J^{\pm}_{1} = J_1 \pm 2S^2B$, and the first three nearest-neighbor vectors are shown in Fig.~\ref{FIG-NiPS3}(b).
For the four-sublattice zigzag phase, $E_{cl} =
N S^2 (J_{1} - 2J_{2} - 3J_{3})/2 + 3N S^4 B/2 + N [A_z \left(\frac{S}{2} \sin^2\theta + S^2 \cos^2\theta\right) + A_x (\frac{S}{2} \cos^2\theta + S^2 \sin^2\theta)]$. 
The corresponding matrices are
\begin{subequations}\label{EQ:ZZBdG}
\begin{eqnarray}\label{EQ:ZZBdGLambda}
\hat{\Lambda}_{\bf q} =
\left(\begin{array}{cccc}
	\mathbb{A}_{\mathbf q}                  & 0      & 0         & \mathbb{D}_{\mathbf q}     \\
    0  & \mathbb{A}_{\mathbf q}                  & \mathbb{D}^{*}_{\mathbf q}     & 0     \\
    0  & \mathbb{D}_{\mathbf q}      & \mathbb{A}_{\mathbf q}                     & 0     \\
    \mathbb{D}^{*}_{\mathbf q}  & 0  & 0     & \mathbb{A}_{\mathbf q}                 \\
\end{array}\right),
\end{eqnarray}
\begin{eqnarray}\label{EQ:ZZBdGDelta}
\hat{\Delta}_{\bf q} =
\left(\begin{array}{cccc}
	\mathbb{E}_{\mathbf q}       & \mathbb{F}_{\mathbf q}     &  \mathbb{G}_{\mathbf q}        & 0     \\
    \mathbb{F}_{-\mathbf q}      & \mathbb{E}_{\mathbf q}     &  0       & \mathbb{G}_{\mathbf q}     \\
    \mathbb{G}_{-\mathbf q}      & 0     &  \mathbb{E}_{\mathbf q}        & \mathbb{F}_{\mathbf q}     \\
    0      & \mathbb{G}_{-\mathbf q}    &  \mathbb{F}_{-\mathbf q}       & \mathbb{E}_{\mathbf q}     \\
\end{array}\right),
\end{eqnarray}
\end{subequations}
with the matrix elements defined as
$\mathbb{A}_{\mathbf q} = A_z (1 - 3\cos^2\theta) + A_x (1 - 3\sin^2\theta) - \left(J_{1} - 2J_{2} - 3J_{3} + 6S^2 B\right) + J_{2} (e^{i{\mathbf q} \boldsymbol\delta^{(2)}_{z}}+e^{-i{\mathbf q} \boldsymbol\delta^{(2)}_{z}})$,
$\mathbb{D}_{\mathbf q} = J^{+}_{1} \sum_{\gamma=x,y} e^{i{\mathbf q} \boldsymbol\delta_{\gamma}}$,
$\mathbb{E}_{\mathbf q} = \frac{1}{2} (A_z \sin^2\theta + A_x \cos^2\theta)$,
$\mathbb{F}_{\mathbf q} = J^{-}_{1} e^{i{\mathbf q}\boldsymbol\delta_{3}} + J_{3}\sum_{\gamma=x,y,z} e^{i{\mathbf q} \boldsymbol\delta^{(3)}_{\gamma}}$,
and $\mathbb{G}_{\mathbf q} = J_{2} \sum_{\gamma=x,y}  \big(e^{i{\mathbf q}\boldsymbol\delta^{(2)}_{\gamma}} + e^{-i{\mathbf q} \boldsymbol\delta^{(2)}_{\gamma}}\big)$.

The quadratic Hamiltonian in Eq.~\eqref{HswZZ} can be diagonalized by a bosonic Bogoliubov transformation $T(\mathbf{q})$. To preserve the canonical commutation relations, the transformation must satisfy the symplectic condition
$T \Sigma T^\dagger = T^\dagger \Sigma T = \Sigma$,
where $\Sigma = \operatorname{diag}(\mathds{1}, -\mathds{1})$. The magnon excitation spectrum is then obtained from the eigenvalues of $\Sigma \hat{\mathbf{H}}_{\mathbf{q}}$, yielding
$\Omega(\mathbf{q}) = \operatorname{diag}\big(\omega_{\mathbf{q},1}, \omega_{\mathbf{q},2}, \cdots, \omega_{\mathbf{q},n_s}\big)$.
For the AFM phase, it is easy to find that
\begin{equation}\label{EQ:DispAFM}
\omega_{\mathbf{q},\pm} = \sqrt{\mathbb{A}^2 - (\mathbb{C} \mp \mathbb{D})^2}.
\end{equation}
For the zigzag phase, the diagonalization procedure is more involved and yields the following magnon dispersions:
\begin{subequations}\label{EQ:DispZZ}
\begin{eqnarray}
\omega_{\mathbf{q},\pm}^{(1)} = \sqrt{u - 2p \pm 2\sqrt{ (q - \mathbb{A} t_+)^2 + v(w - t_+^2) }},
\end{eqnarray}
\begin{eqnarray}
\omega_{\mathbf{q},\pm}^{(2)} = \sqrt{u + 2p \pm 2\sqrt{ (q + \mathbb{A} t_-)^2 + v(w - t_-^2) }}
\end{eqnarray}
\end{subequations}
where $u = \mathbb{A}^2 - \mathbb{E}^2 + |\mathbb{D}|^2 - \mathbb{G}^2 - |\mathbb{F}|^2$,
$v = \mathbb{A}^2 - |\mathbb{F}|^2$,
$w = |\mathbb{D}|^2$,
$p = \mathbb{E}\mathbb{G}$, $q = \operatorname{Re}(\mathbb{F} \mathbb{D}^*)$,
$t_{\pm} = \mathbb{E} \pm \mathbb{G}$.
The resulting spin-wave energy per site, including the zero-point quantum correction, is given by
\begin{equation}
\mathcal{E}_{sw} = \frac{1}{N}\left[E_{cl} + \frac{S}{2} \sum_{\mathbf{q}} \left( \sum_{\upsilon} \omega_{\mathbf{q},\upsilon} - \operatorname{tr}\big( \hat\Lambda_{\mathbf{q}} \big) \right)\right].
\end{equation}

\begin{figure}[!ht]
\centering
\includegraphics[width=0.95\columnwidth, clip]{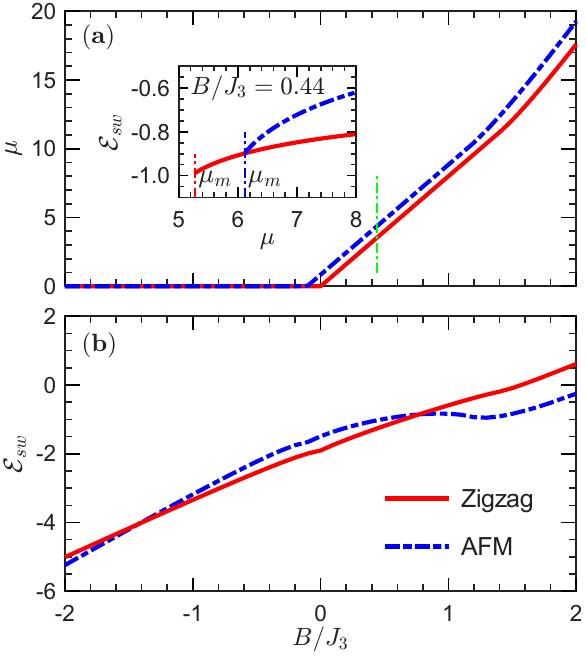}\\
\caption{(a) The minimal $\mu$ as a function of $B/J_3$ for the zigzag phase (red solid line) and AFM phase (blue dot-dashed line) in the $J_1$-$J_3$-$B$ model with $J_1/J_3 = -0.21$. Inset: 
The spin-wave energy of the two phases for the same model at $B/J_3 = 0.44$ as
a function of $\mu$, with the vertical lines denoting the critical $\mu_m$ values at which magnon instabilities are eliminated. (b) Spin-wave energy $\mathcal{E}_{sw}$ evaluated at the corresponding minimal $\mu$ across the same parameter region.}\label{FIG-SWEnergy}
\end{figure}

The spin-wave theory remains valid primarily within its classically ordered regimes. Outside these stable regions, $(\omega_{\mathbf{q},\upsilon})^2$, defined as the eigenvalue of $(\Sigma\hat{\mathbf{H}}_{\mathbf{q}})^2$, may turn negative at certain momenta, indicating the breakdown of spin-wave description. 
The magnon instabilities can be cured by going beyond the harmonic approximation, e.g., by including nonlinear magnon‑magnon interactions or generalized spin‑wave theory. Such approaches renormalize the excitation spectrum and may open a gap in the soft modes through self‑consistent mechanisms or decay processes. However, implementing these methods is technically demanding.
By contrast, this issue can alternatively be addressed by a simple and computationally inexpensive MASWT \cite{Chernyshev2025PRB,Wenzel2012PRL,Coletta2013PRB,Coletta2014PRB}, which introduces a local-field term into the Hamiltonian of the form $\delta\mathcal{H} = \mu \sum_i (S - \mathbf{S}_i \cdot \mathbf{n}_i)$. As $\delta\mathcal{H}$ affects only the diagonal entries of the BdG Hamiltonian, its effect is equivalent to the substitution $\mathbb{A}_{\mathbf{q}} \to \mathbb{A}_{\mathbf{q}} + \mu$ in the BdG formulation. The minimal positive value of $\mu$ is determined by requiring that all eigenvalues $(\omega_{\mathbf{q},\upsilon})^2$ of the modified Hamiltonian remain positive definite across the entire Brillouin zone. To this end, we apply the Gershgorin circle theorem to the BdG Hamiltonian to derive an upper bound $\mu_{\mathrm{max}}$ for the parameter $\mu$. The smallest $\mu$ that guarantees positive definiteness of all BdG eigenvalues is then obtained via a bisection search within the interval $[0, \mu_{\mathrm{max}}]$.

Turning to Fig.~\ref{FIG-J1J2J3CPD}(a), we now investigate the effective $J_1$-$J_3$-$B$ model relevant to NiPS$_3$. Since the competing FM and spiral phases lie far away from the extracted parameters, it suffices to focus on the competition between the AFM and zigzag phases.
To elucidate the application of the MASWT, we first examine the representative case with $J_1/J_3 = -0.21$. The approach involves two successive steps: the determination of the minimal augmentation parameter $\mu_m$, followed by the evaluation of the spin-wave energy.
Figure~\ref{FIG-SWEnergy}(a) presents the minimal $\mu_m$ for the AFM and zigzag phases as a function of $B/J_3$. As $B$ increases from negative to positive values, $\mu_m$ is observed to gradually increase from zero. Specifically, for negative biquadratic interaction, the excitation gap of the lowest-energy band closes at the $\bm{\Gamma}$ point, leading to $\mu_m = 0$. In contrast, a finite $\mu_m$ becomes necessary when the biquadratic interaction is positive.
The inset displays the spin-wave energy $\mathcal{E}_{sw}$ of both phases as a function of $\mu$ at $B/J_3 = 0.44$, obtained within the MASWT. Notably, $\mathcal{E}_{sw}$ increases monotonically for $\mu \geq \mu_m$, indicating that the energy at $\mu = \mu_m$ represents a lower bound. The dependence of $\mu_m$ on $B$ is piecewise linear. For the AFM phase, we obtain
\begin{equation}
\mu_m^{AFM} = 
\begin{cases} 
0, & B < B_{s,1} \\
8S^2B - 4J_1, & B_{s,1}\!< B \!< B_{s,2} \\
12S^2B - 6(J_1 + J_3), & B > B_{s,2}
\end{cases},
\end{equation}
with critical fields $B_{s,1} = J_1/(2S^2)$ and $B_{s,2} = (3J_3 + J_1)/(2S^2)$.
For the zigzag phase,
\begin{equation}
\mu_m^{ZZ} = 
\begin{cases} 
0, & B < B_{s,1} \\
8S^2B, & B_{s,1} \!< B \!< B_{s,2} \\
12S^2B + 2(J_1 - 3J_3), & B > B_{s,2}
\end{cases},
\end{equation}
where $B_{s,1} = 0$ and $B_{s,2} = (3J_3 - J_1)/(2S^2)$.
We note, however, that near $B_{s,2}$ the location of the lowest excitation gap becomes somewhat incommensurate, and the actual $\mu_m$ is slightly larger than the above expressions suggest.

\begin{figure}[!ht]
\centering
\includegraphics[width=0.95\columnwidth, clip]{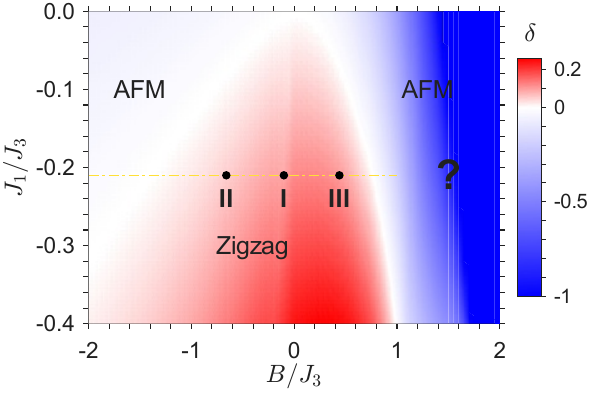}\\
\caption{Intensity plot of $\delta$ [see Eq.~\eqref{EQ:EgDelta}] in the $(B/J_3, J_1/J_3)$ plane for the $J_1$-$J_3$-$B$ model, computed via MASWT. The red (blue) regions indicate parameters favoring the zigzag (AFM) phase, as quantified by the colorbar. Note that energies in the positive $B$ region may carry substantial uncertainty due to innate magnon instabilities.}\label{FIG-SWPhaseDiag}
\end{figure}

With the minimal $\mu_m$ thus determined, the spin-wave energy of both phases is shown in Fig.~\ref{FIG-SWEnergy}(b). The energies of the two magnetic states are found to be highly competitive. A phase transition from the AFM to the zigzag phase occurs at $B_{t,1} \approx -1.4225$, where the two dispersion relations cross. A kink is observed in the energy at $B_{s,1}$, followed by a second crossing at $B_{t,2} \approx 0.7680$. However, a genuine phase transition in this region appears unlikely, as the energy of the AFM phase is significantly underestimated. Despite this limitation, the MASWT has the merit of favoring the zigzag phase as the ground state, at least for moderate positive biquadratic interactions.

We now aim to characterize the full phase competition across the parameter space. To this end, we introduce a normalized energy difference ratio,
\begin{eqnarray}\label{EQ:EgDelta}
\delta = \frac{\mathcal{E}_{sw}^{AFM} - \mathcal{E}_{sw}^{ZZ}}{|\mathcal{E}_{sw}^{AFM}| + |\mathcal{E}_{sw}^{ZZ}|}
\end{eqnarray}
which quantifies the relative energy difference between the two competing magnetic orders. By construction, the zigzag (AFM) phase is energetically favored when $\delta > 0$ ($\delta < 0$).

As shown in Fig.~\ref{FIG-SWPhaseDiag}, for $B/J_3 < 0$, the small magnitude of $|\delta|$-indicated by the pale coloring-reflects strong competition between the two phases and suggests the vicinity of a phase transition. In contrast, while a distinct line of $\delta = 0$ is clearly visible on the positive side of $B/J_3$, the interpretation of a genuine phase transition in this region remains less certain.
Notwithstanding the potential bias inherent in the MASWT approach, the calculated energies consistently favor the zigzag ordering across all three data sets considered, as displayed in Table~\ref{TAB:NiPS3}. Although all three parameter sets stabilize the zigzag phase at the spin-wave level, their discrimination necessitates unbiased many-body calculations that transcend this approximate framework.

%%============================================================================
\subsection{Magnetic order parameters}

Having established that all three data sets favor the zigzag order at the spin-wave level, a crucial question naturally arises: does this magnetic ordering persist in the presence of quantum fluctuations? To address this issue, we performed a series of DMRG calculations on a 24-site hexagonal cluster \cite{White1992PRL,Peschel1999,Schollwock2005RMP}, which is compatible with both zigzag and AFM orders. As a diagnostic tool, we compute the static spin structure factor (SSSF), defined as
\begin{equation}
\mathbb{S}_N^{\alpha\beta}({\bf{Q}})=\frac{1}{N}\sum_{ij}
    \langle{S^{\alpha}_i {S^{\beta}_j}}\rangle e^{i{\bf{Q}}\cdot{({\bm{R}}_i-{\bm{R}}_j)}}.
\end{equation}
where ${\bm{R}}_i$ denotes the position of spin $i$, and ${\bf{Q}}$ is the ordering wavevector.

\begin{figure}[!ht]
\centering
\includegraphics[width=0.95\columnwidth, clip]{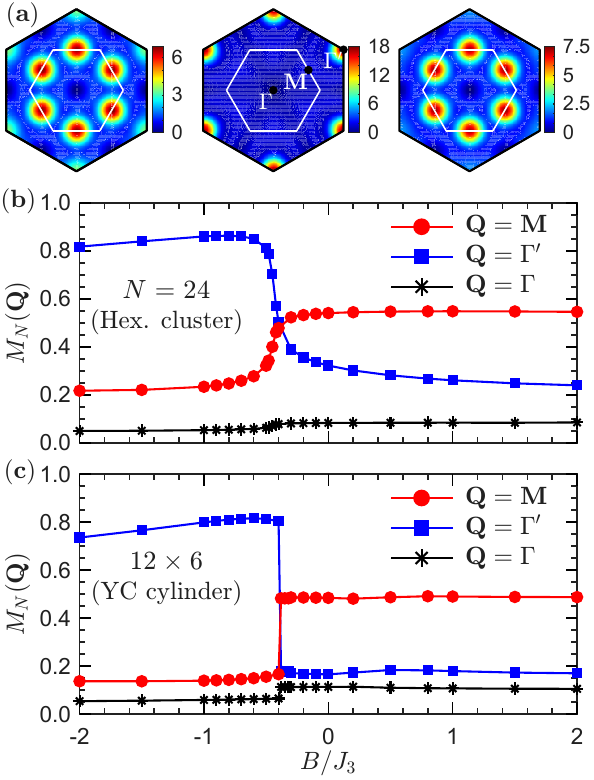}\\
\caption{(a) SSSF for data sets I, II, and III shown in Table~\ref{TAB:NiPS3} (from left to right). (b) The magnetic order parameters $M_N(\mathbf{Q})$ for $\mathbf{Q} = \mathbf{M}$ (red circles), $\boldsymbol{\Gamma}'$ (blue squares), and $\boldsymbol{\Gamma}$ (black asterisks) in the $J_1$-$J_3$-$B$ model at $J_1/J_3 = -0.21$ on a 24-site hexagonal cluster. (c) The same as (b) but for a $12\times6$ YC cylinder.}\label{FIG-DMRG}
\end{figure}

The SSSFs for the three representative data sets are shown in Fig.~\ref{FIG-DMRG}(a). For data set I ($B/J_3 \approx -0.11$) and data set III ($B/J_3 \approx 0.44$), the structure factor exhibits a pronounced peak at the $\textbf{M}$ point, signaling the formalization of zigzag magnetic order. In contrast, data set II ($B/J_3 \approx -0.60$) displays a dominant peak at the $\boldsymbol{\Gamma}'$ point, indicating that the ground state is instead AFM ordered.
To quantitatively characterize the $B$-driven quantum phase transition, we evaluated the magnetic order parameter associated with each ordering wavevector, defined as $M_N({\bf{Q}}) = \sqrt{\mathbb{S}_N({\bf{Q}})/N}$, where $\mathbb{S}_N({\bf{Q}}) = \sum_{\alpha}\mathbb{S}_N^{\alpha\alpha}({\bf{Q}})$. We note that $\mathbb{S}_N({\bf{Q}})$ is averaged over symmetry-equivalent high-symmetry point. Figure~\ref{FIG-DMRG}(b) presents $M_N({\bf{Q}})$ for ${\bf{Q}} = \textbf{M}$ (red circles) and ${\bf{Q}} = \boldsymbol{\Gamma}'$ (blue squares), along with ${\bf{Q}} = \boldsymbol{\Gamma}$ (black asterisks) for reference. The results reveal a clear AFM-zigzag phase transition that occurs at $B/J_3 \simeq -0.44(1)$ within the interval $-2 \leq B/J_3 \leq 2$, while no additional transitions are observed elsewhere in this range. 

To verify the robustness of our DMRG results, we performed additional calculations on a larger $12\times6$ \textrm{YC} cylinder (referring to a cylindrical cluster of size $L_x \times L_y$ with periodic boundary conditions along the $y$ direction and open boundaries along $x$ axis). On this geometry, we again observe a direct phase transition from the AFM to the zigzag order, with no intermediate incommensurate or competing phase, see Fig.~\ref{FIG-DMRG}(c). The extracted transition point $B/J_3 \simeq -0.39(1)$ is very close to that obtained on the $24$-site cluster, confirming that finite-size effects are under control. These findings demonstrate that our conclusions are robust against both system size and boundary condition.
Therefore, our DMRG results indicate that a positive biquadratic interaction provides a more pronounced stabilization of the zigzag magnetic order compared to its negative counterpart.

%%%%%%%%%%%%%%%%%%%%%%%%%%%%%%%%%%%%%%%%%%%%%%%%%%%%%%%%%%%%%%%%%%%%%%%%%%%%%%
\section{Conclusions}

In summary, we have constructed a comprehensive spin Hamiltonian for NiPS$_3$ encompassing exchange couplings and single-ion anisotropy. DFT calculations point to a minimal $J_1$-$J_3$-$B$ model wherein the third-nearest-neighbor Heisenberg exchange $J_3 >0$ constitutes the dominant interaction. While the ratio $J_1/J_3 \approx -0.21$ aligns with previously reported values, our most striking finding is a positive biquadratic coupling with $B/J_3 \approx 0.44$,  in contrast to prior proposals that suggested a negative $B$.
Although the sign of the biquadratic interaction does not induce a phase transition at the classical level, conventional spin-wave theory encounters instabilities when the biquadratic term assumes positive values. To circumvent this issue, we employ MASWT, which introduces an effective chemical potential into the Hamiltonian, thereby lifting the energy levels and eliminating magnon instabilities. A comparative analysis of spin-wave energies reveals that the zigzag phase is energetically favored over the competing AFM phase. For fixed $J_1/J_3 = -0.21$, we identify an AFM-to-zigzag phase transition occurring at a critical value of $B_t/J_3 \simeq -0.44(1)$ (for 24-site hexagonal cluster) or $B_t/J_3 \simeq -0.39(1)$ (for $12 \times 6$ \textrm{YC} cylinder), demonstrating that once quantum fluctuations are taken into account, a positive biquadratic interaction serves to stabilize the zigzag ground state.

The positive biquadratic exchange identified here stands in stark contrast to prevailing negative-$B$ proposals. Notably, conventional spin-wave theory remains applicable when $B<0$, whereas positive $B$ induces magnon instabilities that necessitate the MASWT employed in this work. In the proposed model, single-ion anisotropies further pin the zigzag moments and determine the low-energy magnon gaps.
Experimental fingerprints of NiPS$_3$, including out-of-plane moment canting \cite{Wildes2015PRB,Lane2020PRB,Lee2024AFM}, multiple low-energy gaps at the zone center \cite{Afanasiev2021SA,Mehlawat2022PRB}, and a bulk zigzag phase to few-layer $\mathbb{Z}_3$ vestigial Potts nematic transition \cite{Sun2024NP} may offer valuable constraints on the value of $B$ as well as single-ion anisotropies. Quantitative comparisons of these features against theoretical predictions under competing assumptions could help resolve the ambiguity.
Further experimental techniques sensitive to four-spin interactions, including inelastic neutron scattering and Raman spectroscopy under magnetic fields, may provide additional constraints on the sign and magnitude of $B$, offering pathways toward a definitive spin Hamiltonian for NiPS$_3$.

\begin{acknowledgements}
We thank C. Huang, L.-C. Zhang, and J. Zhao for useful discussions and support.
This work is supported by the National Natural Science Foundation of China (Grants No. 12304176 and No. 12247183),
the Beijing National Laboratory for Condensed Matter Physics (Grant No. 2025BNLCMPKF022), and the Natural Science Foundation of Jiangsu Province (Grant No. BK20220876).
The computations are partially supported by High Performance Computing Platform of Nanjing University of Aeronautics and Astronautics. 
We also gratefully acknowledge HZWTECH for providing computational facilities.
\end{acknowledgements}

\section*{DATA AVAILABILITY}
The data that support the findings of this article are openly available \cite{DataAndCode}.

%%%%%%%%%%%%%%%%%%%%%%%%%%%%%%%%%%%%%%%%%%%%%%%%%%%%%%%%%%%%%%%%%%%%%%%%%%%%%%%%%%%%%%%%%%%%%%%%%%%%%%%%%

%
%%%%%%%%%%%%%%%%%%%%%%%%%%%%%%%%%%%%%%%%%%%%%%%%%%%%%%%%%%%%%%%%%%%%%%%%%%

%%%%%%%%%%%%%%%%%%%%%%%%%%%%%%%%%%%%%%%%%%%%%%%%%%%%%%%%%%%%%%%%%%%%%%%%%%%%%%%%%%%%%%%%%%%%%%%%%%%%%%%%%%%%%%%%%%%%%%%%%%%%%%%%%%%%%%%%%%%%%%%%

%%%%%%%%%% Merge with supplemental materials %%%%%%%%%%
%% ref: https://arxiv.org/abs/1709.10096

%%%%%%%%%%%%%%%%%%%%%%%%%%%%%%%%%%%%%%%%%%%%%%%%%%%%%%%%%%%%%%%%%%%%%%%%%%%%%%%%%%%%%%%%%%%%%%%%%%%%%%%%%%%%%%%%%%%%%%%%%%%%%%%%%%%%%%%%%%%%%%%%

\clearpage

\onecolumngrid

%%%%%%%%%% Merge with supplemental materials %%%%%%%%%%
%%%%%%%%%% Prefix a "S" to all equations, figures, tables and reset the counter %%%%%%%%%%
\newpage

\newcounter{sectionSM}
\newcounter{equationSM}
\newcounter{figureSM}
\newcounter{tableSM}
\stepcounter{equationSM}
\setcounter{section}{0}
\setcounter{equation}{0}
\setcounter{figure}{0}
\setcounter{table}{0}
\setcounter{page}{1}
\makeatletter
\renewcommand{\thesection}{\textsc{S}\arabic{section}}
\renewcommand{\theequation}{\textsc{S}\arabic{equation}}
\renewcommand{\thefigure}{\textsc{S}\arabic{figure}}
\renewcommand{\thetable}{\textsc{S}\arabic{table}}

% \onecolumngrid

%%%%%%%%%%%%%%%%%%%%%%%%%%%%%
\begin{center}
{\large{\bf Supplemental Material for\\
``Stabilization of zigzag order in NiPS$_3$ via positive biquadratic interaction''}}
\end{center}
\begin{center}
Qiang Luo,$^{1,\;2}$ Shuhang Yang,$^{1}$ Xiaoying Wang,$^{1}$ Mengdong Li,$^{1}$ Zhengyu Jiang,$^{1}$ Chunlan Ma,$^{3}$ and Yan Zhu$^{1,\;2}$ \\
\quad\\
$^1$\textit{College of Physics, Nanjing University of Aeronautics and Astronautics, Nanjing, 211106, China}\\
$^2$\textit{Key Laboratory of Aerospace Information Materials and Physics (NUAA), MIIT, Nanjing, 211106, China}\\
$^3$\textit{School of Physical Science and Technology, Suzhou University of Science and Technology, Suzhou 215009, China}\\
(Dated: July 18, 2026)
\quad\\
\end{center}

%\begin{widetext}
%\end{widetext}

% \twocolumngrid
\onecolumngrid

%%%%%%%%%%%%%%%%%%%%%%%%%%%%%
In this Supplemental Material, we present the numerical details of exchange couplings derived from different combinations of magnetic configurations, as well as the derivation of the Bogoliubov–de Gennes (BdG) Hamiltonians for the AFM and zigzag phases.

\vspace{-0.00cm}
\section{Extraction of Exchange Couplings}\label{SMSec:I}
\subsection{GGA+$U$ method with spin-orbit coupling with $U_{\text{eff}} = 4.0$ eV}\label{SMSec:IA}

\begin{figure}[!ht]
\centering
\includegraphics[width=0.60\columnwidth, clip]{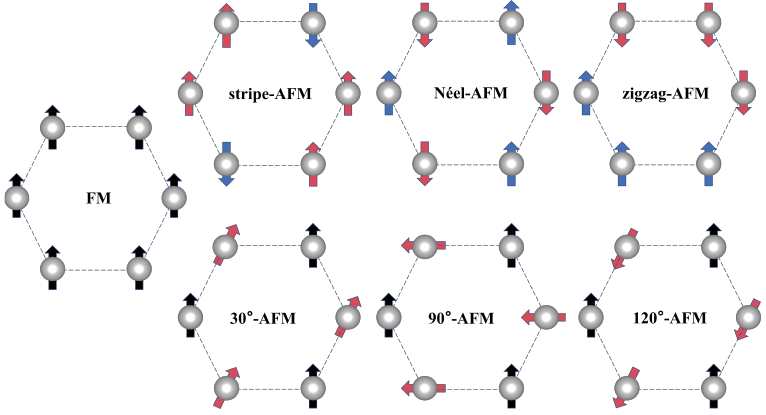}\\
\caption{Schematic illustrations of various linear and nonlinear magnetic configurations.}\label{FIGSM-SpinConfig}
\end{figure}

To extract the exchange couplings $J_1$, $J_2$, $J_3$, and $B$ in the proposed Hamiltonian [Eq.~(\textcolor{red}{1}) in the main text], we consider the FM, N\'eel-type AFM, stripe-AFM, and zigzag-AFM configurations, along with three noncollinear states with relative spin angles of $30^\circ$, $90^\circ$, and $120^\circ$, as illustrated in Fig.~\ref{FIGSM-SpinConfig}.

Using the four-state energy mapping method with spin-orbit coupling, the energy differences relative to the FM state are given by
\begin{alignat}{2}
-12J_1 - 0J_2 - 12J_3 - 0B_1 &= E_{\text{N\'eel-AFM}} - E_{\text{FM}} &&= -136.158~\text{meV}, \\
-8J_1 - 16J_2 - 0J_3 - 0B_1 &= E_{\text{stripe-AFM}} - E_{\text{FM}} &&= \phantom{-}30.135~\text{meV}, \\
-4J_1 - 16J_2 - 12J_3 - 0B_1 &= E_{\text{zigzag-AFM}} - E_{\text{FM}} &&= -157.365~\text{meV}, \\
-9J_1 - 0J_2 - 9J_3 - \frac{9}{2}B_1 &= E_{\text{120-AFM}} - E_{\text{FM}} &&= -135.486~\text{meV}, \\
-6J_1 - 0J_2 - 6J_3 - 6B_1 &= E_{\text{90-AFM}} - E_{\text{FM}} &&= -105.354~\text{meV}, \\
-(6 - 3\sqrt{3})J_1 - 0J_2 - (6 - 3\sqrt{3})J_3 - \frac{3}{2}B_1 &= E_{\text{30-AFM}} - E_{\text{FM}} &&= -18.305~\text{meV}.
\end{alignat}
The resulting fitted parameters are $(J_1, J_2, J_3, B) \approx (-3.1309, -0.3182, 14.5817, 6.4868)$~meV.

To assess the precision of the estimates, we consider various combinations of these spin configurations. For brevity, the configurations are abbreviated as summarized in Table~\ref{TAB:NameConfig}. Mathematically, at least four different spin configurations should be included to estimate the four exchange parameters. However, four of the six configurations happen to share identical coefficients for $J_1$ and $J_3$. This can be seen clearer by rewritten Eqs.~(S1)-(S6) as
\begin{align}
\left\{
  \begin{array}{clcl}
  \rm{M1}: & 1 \cdot J_1 + 0 \cdot J_2 + 1 \cdot J_3 + 0 \cdot B &=& \phantom{+} 11.347 ~\text{meV} \\
  \rm{M2}: & 1 \cdot J_1 + 2 \cdot J_2 + 0 \cdot J_3 + 0 \cdot B &=& -3.767 ~\text{meV} \\
  \rm{M3}: & 1 \cdot J_1 + 4 \cdot J_2 + 3 \cdot J_3 + 0 \cdot B &=& \phantom{+} 39.341 ~\text{meV} \\
  \rm{M4}: & 1 \cdot J_1 + 0 \cdot J_2 + 1 \cdot J_3 + \frac{1}{2} \cdot B &=& \phantom{+} 15.054 ~\text{meV} \\
  \rm{M5}: & 1 \cdot J_1 + 0 \cdot J_2 + 1 \cdot J_3 + 1 \cdot B &=& \phantom{+} 17.559 ~\text{meV} \\
  \rm{M6}: & 1 \cdot J_1 + 0 \cdot J_2 + 1 \cdot J_3 + \frac{2+\sqrt3}{2} \cdot B &=& \phantom{+} 22.772 ~\text{meV} \\
  \end{array}
\right..
\end{align}
Therefore, to avoid underfitting, both the stripe-AFM configuration $\rm M2$ and the zigzag-AFM configuration $\rm M3$ must be included to uniquely resolve all exchange parameters.

\begin{table}[htbp]
\centering
\caption{Magnetic configurations considered in the fitting of the spin Hamiltonian parameters.}
\label{TAB:NameConfig}
\begin{tabular*}{0.50\textwidth}{@{\extracolsep{\fill}} c c @{}}
\toprule
Name & Magnetic configuration \\
\hline
M0   & FM \\
M1   & N\'eel-AFM \\
M2   & stripe-AFM \\
M3   & zigzag-AFM \\
M4   & Non-collinear 120$^\circ$-AFM \\
M5   & Non-collinear 90$^\circ$-AFM \\
M6   & Non-collinear 30$^\circ$-AFM \\
\hline\hline
\end{tabular*}
\end{table}

The fitting results for all possible combinations of spin configurations are summarized in Table~\ref{TAB:ParaConfig},  where entries that \textbf{include both M2 and M3} and yield estimated $J_3$ values reasonably close to 14.5817 meV are highlighted in bold. These selected results are used to compute the mean and standard error for each exchange coupling, as presented in Fig.~\textcolor{red}{3} of the main text.

\begin{table}[htbp]
\centering
\caption{Extracted Heisenberg exchange parameters ($J$) and biquadratic parameters ($B$) for various combinations of collinear and noncollinear magnetic states. All values are in meV.}
\label{TAB:ParaConfig}
\begin{tabular*}{0.9\textwidth}{@{\extracolsep{\fill}} l c c c c @{}}
\toprule
Combination & $J_1$ & $J_2$ & $J_3$ & $B_1$ \\
\hline
\bfseries M1, M2, M3, M4, M5, M6 & \bfseries -3.1309 & \bfseries -0.3182 & \bfseries 14.5817 & \bfseries 6.4868 \\
\bfseries M2, M3, M4, M5, M6    & \bfseries -2.3507 & \bfseries -0.7084 & \bfseries 14.8418 & \bfseries 5.1049 \\
M1, M3, M4, M5, M6    & 4.8121 & 3.6533 & 6.6387 & 6.4868 \\
M1, M2, M4, M5, M6    & 4.6706 & -4.2190 & 6.7801 & 6.4868 \\
\bfseries M1, M2, M3, M5, M6    & \bfseries -3.2087 & \bfseries -0.2794 & \bfseries 14.5558 & \bfseries 6.2066 \\
\bfseries M1, M2, M3, M4, M6    & \bfseries -3.1898 & \bfseries -0.2888 & \bfseries 14.5621 & \bfseries 7.2379 \\
\bfseries M1, M2, M3, M4, M5    & \bfseries -3.1347 & \bfseries -0.3164 & \bfseries 14.5805 & \bfseries 6.5104 \\
\bfseries M1, M2, M3, M4       & \bfseries -3.2091 & \bfseries -0.2791 & \bfseries 14.5556 & \bfseries 7.4150 \\
\bfseries M1, M2, M3, M5       & \bfseries -3.2091 & \bfseries -0.2791 & \bfseries 14.5556 & \bfseries 6.2125 \\
\bfseries M1, M2, M3, M6       & \bfseries -3.2091 & \bfseries -0.2791 & \bfseries 14.5556 & \bfseries 6.1227 \\
M1, M2, M4, M5       & 4.6684 & -4.2179 & 6.7774 & 6.5104 \\
M1, M2, M4, M6       & 4.6358 & -4.2016 & 6.7366 & 7.2379 \\
M1, M2, M5, M6       & 4.6246 & -4.1960 & 6.7226 & 6.2066 \\
M1, M3, M4, M5       & 4.8090 & 3.6555 & 6.6368 & 6.5104 \\
M1, M3, M4, M6       & 4.7641 & 3.6881 & 6.6082 & 7.2379 \\
M1, M3, M5, M6       & 4.7487 & 3.6993 & 6.5984 & 6.2066 \\
M1, M4, M5, M6       & 5.7254 & 0 & 5.7254 & 6.4868 \\
\bfseries M2, M3, M4, M5       & \bfseries -2.3073 & \bfseries -0.7301 & \bfseries 14.8563 & \bfseries 5.0100 \\
\bfseries M2, M3, M4, M6       & \bfseries -2.5472 & \bfseries -0.6101 & \bfseries 14.7763 & \bfseries 5.6497 \\
\bfseries M2, M3, M5, M6       & \bfseries -3.0641 & \bfseries -0.3517 & \bfseries 14.6040 & \bfseries 6.0191 \\
M2, M4, M5, M6       & 5.1330 & -4.4502 & 7.3581 & 5.1049 \\
M3, M4, M5, M6       & 5.4479 & 3.1909 & 7.0433 & 5.1049 \\
\hline\hline
\end{tabular*}
\end{table}

\subsection{GGA+$U$ method without spin-orbit coupling with $U_{\text{eff}} = 4.0$ eV}\label{SMSec:IB_noSOC}

To assess the influence of SOC, we performed additional calculations under the same conditions ($U_{\mathrm{eff}} = 4$~eV) but omitting SOC. 
The energy differences relative to the FM state and the exchange parameters are presented in Tables~\ref{TabSM:Ueff4Energy} and~\ref{TabSM:Ueff4Coupling}, respectively. 
These results confirm that excluding SOC does not affect the main conclusion, namely that the parameter $B$ remains positive with $B/J_3 \approx 0.44$.

\begin{table}[htbp]
\centering
\caption{Magnetic exchange energies (in meV) with and without SOC.}
\label{TabSM:Ueff4Energy}
\begin{tabular*}{0.90\textwidth}{@{\extracolsep{\fill}} lccccccc @{}}
%\begin{tabular}{lccccccc}
\toprule
SOC & $E_{\rm{Neel-AFM}}$ & $E_{\rm{stripe-AFM}}$ & $E_{\rm{zigzag-AFM}}$ & $E_{\rm{120^{\circ}-AFM}}$ &  $E_{\rm{90^{\circ}-AFM}}$ &  $E_{\rm{30^{\circ}-AFM}}$ \\
\hline
With  & $-136.16$ & $30.14$ & $-157.37$ & $-135.49$ & $-105.35$ & $-18.31$ \\
Without & $-136.97$ & $30.18$ & $-158.20$ & $-113.81$ & $-116.40$ & $-16.34$ \\
\hline\hline
\end{tabular*}
\end{table}

\begin{table}[th!]
\centering
\caption{Extracted magnetic parameters (in meV) and their ratios with and without SOC.}
\label{TabSM:Ueff4Coupling}
\begin{tabular*}{0.60\textwidth}{@{\extracolsep{\fill}} lccccccc @{}}
\toprule
SOC & $J_1$ & $J_2$ & $J_3$ & $B_1$ & $J_1/J_3$ & $J_2/J_3$ & $B_1/J_3$ \\
\hline
With  & $-3.13$ & $-0.32$ & $14.58$ & $6.49$ & $-0.21$ & $-0.02$ & $0.44$ \\
Without & $-3.54$ & $-0.12$ & $14.52$ & $6.53$ & $-0.24$ & $-0.01$ & $0.45$ \\
\hline\hline
\end{tabular*}
\end{table}

\subsection{GGA+$U$ method with spin-orbit coupling at different $U_{\text{eff}}$}\label{SMSec:IC_Ueff}

To examine the robustness of our results with respect to the Hubbard $U_{\mathrm{eff}}$, we performed additional DFT calculations using $U_{\mathrm{eff}} = 3$~eV and $5$~eV and compared them with the original results obtained at $U_{\mathrm{eff}} = 4$~eV. 
The energy differences relative to the FM state and the exchange parameters at different $U_{\mathrm{eff}}$ are presented in Tables~\ref{TabSM:Ueff345Energy} and~\ref{TabSM:Ueff345Coupling}, respectively. 
The results clearly show that, as $U_{\mathrm{eff}}$ increases, the absolute values of all interaction parameters decrease monotonically. Importantly, however, the ratios between the parameters remain within a narrow range. In particular, $B$ is consistently positive for all three $U_{\mathrm{eff}}$ values, and the ratio $B/J_3$ is stable at $0.44^{+0.03}_{-0.02}$. This demonstrates that our extracted parameters---and thus our main physical conclusions---are robust against variations in the Hubbard $U_{\mathrm{eff}}$.

\begin{table}[th!]
\centering
\caption{Magnetic exchange energies (in meV) at different $U_{\text{eff}}$.}
\label{TabSM:Ueff345Energy}
\begin{tabular*}{0.90\textwidth}{@{\extracolsep{\fill}} cccccccc @{}}
%\begin{tabular}{lccccccc}
\toprule
$U_{\text{eff}}$ (eV) & $E_{\rm{Neel-AFM}}$ & $E_{\rm{stripe-AFM}}$ & $E_{\rm{zigzag-AFM}}$ & $E_{\rm{120^{\circ}-AFM}}$ &  $E_{\rm{90^{\circ}-AFM}}$ &  $E_{\rm{30^{\circ}-AFM}}$ \\
\hline
3 & $-178.83$ & $35.34$ & $-204.08$ & $-178.73$ & $-133.10$ & $-24.16$ \\
4 & $-136.16$ & $30.14$ & $-157.37$ & $-135.49$ & $-105.35$ & $-18.31$ \\
5 & $-103.51$ & $24.87$ & $-120.44$ & $-101.91$ & $-82.32$ & $-17.13$ \\
\hline\hline
\end{tabular*}
\end{table}

\begin{table}[th!]
\centering
\caption{Extracted magnetic parameters (in meV) and their ratios at different $U_{\text{eff}}$.}
\label{TabSM:Ueff345Coupling}
\begin{tabular*}{0.60\textwidth}{@{\extracolsep{\fill}} cccccccc @{}}
\toprule
$U_{\text{eff}}$ (eV) & $J_1$ & $J_2$ & $J_3$ & $B_1$ & $J_1/J_3$ & $J_2/J_3$ & $B_1/J_3$ \\
\hline
3  & $-3.62$ & $-0.40$ & $18.74$ & $7.94$ & $-0.19$ & $-0.021$ & $0.42$ \\
4  & $-3.13$ & $-0.32$ & $14.58$ & $6.49$ & $-0.21$ & $-0.022$ & $0.44$ \\
5  & $-2.61$ & $-0.25$ & $11.24$ & $5.25$ & $-0.23$ & $-0.022$ & $0.47$ \\
\hline\hline
\end{tabular*}
\end{table}

\subsection{Comparison of GGA+$U$ with spin-orbit coupling ($U_{\text{eff}} = 4.0$ eV) and HSE06 functional}\label{SMSec:IB_HSE06}

We carried out calculations using the hybrid HSE06 functional. 
The energy differences relative to the FM state and the exchange parameters extracted with the GGA+$U$ and HSE06 functionals are presented in Tables~\ref{TabSM:MethodEnergy} and~\ref{TabSM:MethodCoupling}, respectively. 
The HSE06 results further confirm that the parameter $B$ remains positive and, importantly, its magnitude is largely unchanged compared to the GGA+$U$ results. 
Among the exchange parameters, $J_3$ is the most notably affected by HSE06, likely because its hybrid functional treatment improves both Ni--$3d$ and S--$3p$ states, which tends to moderate the long-range $p$--$d$ hopping along the Ni--S--S--Ni super-superexchange pathway that primarily governs $J_3$ \cite{SMFranchini2007PRB}. Although $J_3$ is somewhat reduced under HSE06, it remains the dominant interaction. Importantly, despite this moderate variation, the sign and magnitude of $B$ are found to be remarkably consistent between the two functionals, which reinforces the robustness of our key findings.

\begin{table}[th!]
\centering
\caption{Magnetic exchange energies (in meV) by different methods.}
\label{TabSM:MethodEnergy}
\begin{tabular*}{0.90\textwidth}{@{\extracolsep{\fill}} cccccccc @{}}
%\begin{tabular}{lccccccc}
\toprule
$U_{\text{eff}}$ (eV) & $E_{\rm{Neel-AFM}}$ & $E_{\rm{stripe-AFM}}$ & $E_{\rm{zigzag-AFM}}$ & $E_{\rm{120^{\circ}-AFM}}$ &  $E_{\rm{90^{\circ}-AFM}}$ &  $E_{\rm{30^{\circ}-AFM}}$ \\
\hline
GGA+$U$ & $-136.16$ & $30.14$ & $-157.37$ & $-135.49$ & $-105.35$ & $-18.31$ \\
HSE06   & $-77.72$  & $32.39$  & $-97.43$   & ---        & $-78.00$   & --- \\
\hline\hline
\end{tabular*}
\end{table}

\begin{table}[th!]
\centering
\caption{Extracted magnetic parameters (in meV) and their ratios by different methods.}
\label{TabSM:MethodCoupling}
\begin{tabular*}{0.60\textwidth}{@{\extracolsep{\fill}} cccccccc @{}}
\toprule
$U_{\text{eff}}$ (eV) & $J_1$ & $J_2$ & $J_3$ & $B_1$ & $J_1/J_3$ & $J_2/J_3$ & $B_1/J_3$ \\
\hline
GGA+$U$  & $-3.13$ & $-0.32$ & $14.58$ & $6.49$ & $-0.21$ & $-0.022$ & $0.44$ \\
HSE06    & $-3.26$ & $-0.40$ & $9.73$  & $6.52$ & $-0.33$ & $-0.04$  & $0.67$ \\
\hline\hline
\end{tabular*}
\end{table}

\vspace{-0.00cm}
\section{Derivation of the BdG Hamiltonians}\label{SMSec:II}

In the spin-wave theory, each spin operator $\mathbf S_i$ is replaced by bosonic operators $a_i$ and $a_i^\dagger$, which satisfy the canonical commutation relations $[a_i,a_j^\dagger] = \delta_{ij}$ and $[a_i,a_j] = [a_i^\dagger,a_j^\dagger] = 0$. We employ a Holstein-Primakoff decomposition

\begin{eqnarray}
	\tilde{S}_i^+ = \sqrt{2S} a_i, \quad
	\tilde{S}_i^- = \sqrt{2S} a_i^\dagger, \quad
	\tilde{S}_i^z = S - a_i^\dagger a_i,
\end{eqnarray}
or equivalently,
\begin{eqnarray}
	\tilde{S}_i^x = \sqrt{\frac{S}{2}} (a_i + a_i^\dagger), \quad
    \tilde{S}_i^x = -\imath\sqrt{\frac{S}{2}} (a_i - a_i^\dagger), \quad
	\tilde{S}_i^z = S - a_i^\dagger a_i.
\end{eqnarray}
Here, $\tilde{S}_i^z \equiv (\mathbf S \cdot \mathbf n)$ is the spin component along the classical spin direction $\mathbf n$ and $\tilde{S}_i^\pm = (\mathbf S_i \cdot \mathbf e) \pm [\mathbf S_i \cdot (\mathbf n \times \mathbf e)]$ are the ladder operators consisting of the orthogonal spin components, with $\mathbf e$ being an (arbitrary) unit vector perpendicular to $\mathbf n$ \cite{SMJanssen2019JPCM}.
For any spin that is either parallel ($\tau = 1$) or antiparallel ($\tau = -1$) to the classical moment direction $\mathbf{n}$,
we can rewrite the spin operator as
\begin{align}\label{SpinHP}
\mathbf{S}_{\tau,i}
&= \tilde{S}_i^x \mathbf{e} +\tilde{S}_i^y (\mathbf n \times \mathbf e) + \tilde{S}_i^z \mathbf{n} 
= \sqrt{\frac{S}{2}}(a_i+a_i^{\dagger}){\mathbf{e}} + \tau\sqrt{\frac{S}{2}}(-\ii a_i+\ii a_i^{\dagger}) (\mathbf n \times \mathbf e)
+\tau(S-a_i^{\dagger}a_i) {\mathbf{n}}
\end{align}

The factor $\tau$ encodes the sublattice magnetization direction, allowing for a unified treatment of ferromagnetic ($\tau=1$) and antiferromagnetic ($\tau=-1$) bonds. In the following, we apply this formalism to derive the linearized spin-wave Hamiltonian, retaining only terms up to quadratic order in the bosonic operators. These quadratic terms are subsequently diagonalized via a Bogoliubov transformation to obtain the magnon dispersion relations.

\begin{figure}[!ht]
\centering
\includegraphics[width=0.50\columnwidth, clip]{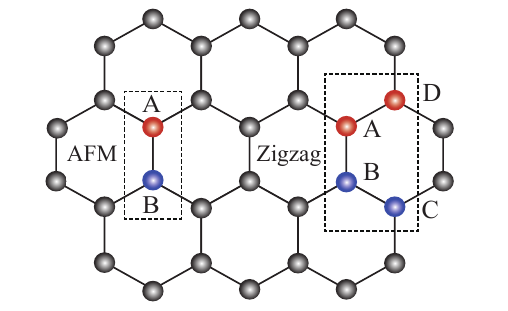}\\
\caption{Unit cells of two-sublattice AFM phase and four-sublattice zigzag phase.}\label{FIGSM-UnitCell}
\end{figure}

\subsection{Two-sublattice phases: AFM phase as an example}

\subsubsection{The nearest-neighbor interactions $J_1$ and $B$}

By substituting the Holstein-Primakoff representation Eq.~\eqref{SpinHP} into the spin-spin interaction, we obtain an expression in terms of bosonic operators. The result simplifies significantly because the product $\tau\tau'$ determines the nature of the bond. For a bond between sites $i$ and $j$ on different sublattices, we define $\tau_{\gamma} = \tau\tau' = \pm 1$. The linear spin-wave approximation involves keeping terms up to quadratic order in the boson operators, which is valid for large $S$ or low temperatures where the magnon density is small. Using Eq.~\eqref{SpinHP} we find that
\begin{eqnarray}\label{SiSjTiTj}
\mathbf{S}_{\tau,i} \cdot \mathbf{S}_{\tau',j}
&=& \tau\tau'S^2 - \tau\tau'S (a_i^{\dagger}a_i + b_j^{\dagger}b_j) + S\frac{1+\tau\tau'}{2} (a_ib_j^{\dagger} + a_i^{\dagger}b_j) + S\frac{1-\tau\tau'}{2} (a_ib_j + a_i^{\dagger}b_j^{\dagger}).
\end{eqnarray}
The expression only depends on the product of $\tau$ and $\tau'$.
For any ${\langle i j \rangle_\gamma}$ bond,
we define $\tau_{\gamma} = \tau\tau'$ and Eq.~\eqref{SiSjTiTj} turns to be
\begin{eqnarray}\label{SiSjTr}
\mathbf{S}_{\tau,i} \cdot \mathbf{S}_{\tau',j}
&=& \tau_{\gamma}S^2 - \tau_{\gamma} S (a_i^{\dagger}a_i + b_j^{\dagger}b_j) + S\frac{1+\tau_{\gamma}}{2} (a_ib_j^{\dagger} + a_i^{\dagger}b_j) + S\frac{1-\tau_{\gamma}}{2} (a_ib_j + a_i^{\dagger}b_j^{\dagger}).
\end{eqnarray}

For the biquadratic interaction $(\mathbf S_i \cdot \mathbf S_j)^2$, we square the above expression and keep terms up to quadratic order in bosonic operators. This yields

\begin{eqnarray}\label{SiSjSquareTr}
\left(\mathbf{S}_{\tau,i} \cdot \mathbf{S}_{\tau',j}\right)^2
&=& \Big[\tau_{\gamma}S^2 - \tau_{\gamma} S (a_i^{\dagger}a_i + b_j^{\dagger}b_j) + S\frac{1+\tau_{\gamma}}{2} (a_ib_j^{\dagger} + a_i^{\dagger}b_j) + S\frac{1-\tau_{\gamma}}{2} (a_ib_j + a_i^{\dagger}b_j^{\dagger}) \Big]^2  \nonumber \\
&\approx& \left(\tau_{\gamma}S^2\right)^2 + 2\tau_{\gamma}S^3 \Big[- \tau_{\gamma} (a_i^{\dagger}a_i + b_j^{\dagger}b_j) + \frac{1+\tau_{\gamma}}{2} (a_ib_j^{\dagger} + a_i^{\dagger}b_j) + \frac{1-\tau_{\gamma}}{2} (a_ib_j + a_i^{\dagger}b_j^{\dagger}) \Big] \nonumber \\
&=& S^4 + 2S^3 \Big[-(a_i^{\dagger}a_i + b_j^{\dagger}b_j) + \frac{1+\tau_{\gamma}}{2} (a_ib_j^{\dagger} + a_i^{\dagger}b_j) - \frac{1-\tau_{\gamma}}{2} (a_ib_j + a_i^{\dagger}b_j^{\dagger}) \Big]
\end{eqnarray}

We now transform the Heisenberg ($J_1$) interaction into momentum space by introducing Fourier transformed bosonic operators. The summation over bonds yields

\begin{eqnarray}\label{HsbgMom}
\sum_{\langle i j \rangle_\gamma} J_{\gamma} \mathbf S_i \cdot \mathbf S_j
%%%%%%%%%%%%%%%%%%%%%%  line-3
&=&  N_sS^2\sum_{\gamma} J_{\gamma}\tau_{\gamma}
- S\sum_{\gamma} J_{\gamma}\tau_{\gamma} \sum_{\mathbf q} (a_{\mathbf q}^{\dagger}a_{\mathbf q} + b_{\mathbf q}^{\dagger}b_{\mathbf q})  \nonumber \\
&&  + S\sum_{\gamma} J_{\gamma}\frac{1+\tau_{\gamma}}{2}
\sum_{\mathbf q} (e^{i{\mathbf q}\boldsymbol\delta_{\gamma}}a_{\mathbf q}^{\dagger}b_{\mathbf q} + \mathrm{h.c.}) \nonumber \\
&&  + S\sum_{\gamma} J_{\gamma}\frac{1-\tau_{\gamma}}{2}
\sum_{\mathbf q} (e^{i{\mathbf q}\boldsymbol\delta_{\gamma}}a_{\mathbf q}^{\dagger}b_{-\mathbf q}^{\dagger} + \mathrm{h.c.})
\end{eqnarray}

Similarly, the biquadratic interaction in momentum space is given by

\begin{eqnarray}\label{HsbgMom}
\sum_{\langle i j \rangle_\gamma} B_{\gamma} (\mathbf S_i \cdot \mathbf S_j)^2
%%%%%%%%%%%%%%%%%%%%%%  line-3
&=&  N_sS^4\sum_{\gamma} B_{\gamma}
- 2S^3\sum_{\gamma} B_{\gamma} \sum_{\mathbf q} (a_{\mathbf q}^{\dagger}a_{\mathbf q} + b_{\mathbf q}^{\dagger}b_{\mathbf q})  \nonumber \\
&&  + 2S^3\sum_{\gamma} B_{\gamma}\frac{1+\tau_{\gamma}}{2}
\sum_{\mathbf q} (e^{i{\mathbf q}\boldsymbol\delta_{\gamma}}a_{\mathbf q}^{\dagger}b_{\mathbf q} + \mathrm{h.c.}) \nonumber \\
&&  - 2S^3\sum_{\gamma} B_{\gamma}\frac{1-\tau_{\gamma}}{2}
\sum_{\mathbf q} (e^{i{\mathbf q}\boldsymbol\delta_{\gamma}}a_{\mathbf q}^{\dagger}b_{-\mathbf q}^{\dagger} + \mathrm{h.c.})
\end{eqnarray}

Combining both contributions, the total nearest-neighbor Hamiltonian $\mathcal{H}_{1NN}$ can be written compactly as
\begin{eqnarray}\label{HsbgMom}
\sum_{\langle i j \rangle_\gamma} J_{\gamma} \mathbf S_i \cdot \mathbf S_j + \sum_{\langle i j \rangle_\gamma} B_{\gamma} (\mathbf S_i \cdot \mathbf S_j)^2
%%%%%%%%%%%%%%%%%%%%%%  line-1
&=&  N_sS^2\sum_{\gamma} J_{\gamma}\tau_{\gamma} + N_sS^4\sum_{\gamma} B_{\gamma} - S\sum_{\gamma} (J_{\gamma}\tau_{\gamma} + 2S^2B_{\gamma}) \sum_{\mathbf q} (a_{\mathbf q}^{\dagger}a_{\mathbf q} + b_{\mathbf q}^{\dagger}b_{\mathbf q})  \nonumber \\
&&  + S\sum_{\gamma} (J_{\gamma} + 2S^2B_{\gamma})\frac{1+\tau_{\gamma}}{2}
\sum_{\mathbf q} (e^{i{\mathbf q}\boldsymbol\delta_{\gamma}}a_{\mathbf q}^{\dagger}b_{\mathbf q} + \mathrm{h.c.}) \nonumber \\
&&  + S\sum_{\gamma} (J_{\gamma} - 2S^2B_{\gamma})\frac{1-\tau_{\gamma}}{2}
\sum_{\mathbf q} (e^{i{\mathbf q}\boldsymbol\delta_{\gamma}}a_{\mathbf q}^{\dagger}b_{-\mathbf q}^{\dagger} + \mathrm{h.c.})
\end{eqnarray}

\subsubsection{The second nearest-neighbor interaction $J_2$}

We now consider the second-neighbor interactions. For spins on the $\mathcal{A}$-sublattice, the contributions take the form

\begin{eqnarray}
\mathcal{H}_{2NN}^{\mathcal{A}}
%%%%%%%%%%%%%%%%%%%%%%  line-1
&=& N_sS^2\sum_{\gamma=x,y,z} J_{2,\gamma}\tau^{(2)}_{\gamma}   - 2S \sum_{\mathbf q} \sum_{\gamma=x,y,z} J_{2,\gamma}\tau^{(2)}_{\gamma} (a_{\mathbf q}^{\dagger}a_{\mathbf q})  \nonumber \\
%%%%%%%%%%%%%%%%%%%%%%  line-3
&&  + S \sum_{\mathbf q} \sum_{\gamma=x,y,z} J_{2,\gamma}\frac{1+\tau^{(2)}_{\gamma}}{2} e^{i{\mathbf q}\boldsymbol\delta^{(2)}_{\gamma}} (a_{\mathbf q}^{\dagger}a_{\mathbf q}) + \mathrm{h.c.} \nonumber \\
%%%%%%%%%%%%%%%%%%%%%%  line-4
&&  + S \sum_{\mathbf q} \sum_{\gamma=x,y,z} J_{2,\gamma}\frac{1-\tau^{(2)}_{\gamma}}{2} e^{i{\mathbf q}\boldsymbol\delta^{(2)}_{\gamma}} (a_{\mathbf q}^{\dagger}a_{-\mathbf q}^{\dagger}) + \mathrm{h.c.} \nonumber \\
%%%%%%%%%%%%%%%%%%%%%%  line-1
&=& N_sS^2\sum_{\gamma=x,y,z} J_{2,\gamma}\tau^{(2)}_{\gamma}   \nonumber \\
%%%%%%%%%%%%%%%%%%%%%%  line-2
&&  + S \sum_{\mathbf q} \left(\sum_{\gamma=x,y,z} J_{2,\gamma}\frac{1+\tau^{(2)}_{\gamma}}{2} (e^{i{\mathbf q}\boldsymbol\delta^{(2)}_{\gamma}} + e^{-i{\mathbf q}\boldsymbol\delta^{(2)}_{\gamma}}) -2 \sum_{\gamma=x,y,z} J_{2,\gamma}\tau^{(2)}_{\gamma}\right) (a_{\mathbf q}^{\dagger}a_{\mathbf q}) \nonumber \\
%%%%%%%%%%%%%%%%%%%%%%  line-3
&&  + S \sum_{\mathbf q} \sum_{\gamma=x,y,z} J_{2,\gamma}\frac{1-\tau^{(2)}_{\gamma}}{2} e^{i{\mathbf q}\boldsymbol\delta^{(2)}_{\gamma}} (a_{\mathbf q}^{\dagger}a_{-\mathbf q}^{\dagger}) + \mathrm{h.c.}
\end{eqnarray}

Analogously, for the $\mathcal{B}$-sublattice we obtain
\begin{eqnarray}
\mathcal{H}_{2NN}^{\mathcal{B}}
%%%%%%%%%%%%%%%%%%%%%%  line-1
&=& N_sS^2\sum_{\gamma=x,y,z} J_{2,\gamma}\tau^{(2)}_{\gamma} - 2S \sum_{\mathbf q} \sum_{\gamma=x,y,z} J_{2,\gamma}\tau^{(2)}_{\gamma} (b_{\mathbf q}^{\dagger}b_{\mathbf q})  \nonumber \\
%%%%%%%%%%%%%%%%%%%%%%  line-3
&&  + S \sum_{\mathbf q} \sum_{\gamma=x,y,z} J_{2,\gamma}\frac{1+\tau^{(2)}_{\gamma}}{2} e^{i{\mathbf q}\boldsymbol\delta^{(2)}_{\gamma}} (b_{\mathbf q}^{\dagger}b_{\mathbf q}) + \mathrm{h.c.} \nonumber \\
%%%%%%%%%%%%%%%%%%%%%%  line-4
&&  + S \sum_{\mathbf q} \sum_{\gamma=x,y,z} J_{2,\gamma}\frac{1-\tau^{(2)}_{\gamma}}{2} e^{i{\mathbf q}\boldsymbol\delta^{(2)}_{\gamma}} (b_{\mathbf q}^{\dagger}b_{-\mathbf q}^{\dagger}) + \mathrm{h.c.} \nonumber \\
%%%%%%%%%%%%%%%%%%%%%%  line-1
&=& N_sS^2\sum_{\gamma=x,y,z} J_{2,\gamma}\tau^{(2)}_{\gamma}   \nonumber \\
%%%%%%%%%%%%%%%%%%%%%%  line-2
&&  + S \sum_{\mathbf q} \left(\sum_{\gamma=x,y,z} J_{2,\gamma}\frac{1+\tau^{(2)}_{\gamma}}{2} (e^{i{\mathbf q}\boldsymbol\delta^{(2)}_{\gamma}} + e^{-i{\mathbf q}\boldsymbol\delta^{(2)}_{\gamma}}) -2 \sum_{\gamma=x,y,z} J_{2,\gamma}\tau^{(2)}_{\gamma}\right) (b_{\mathbf q}^{\dagger}b_{\mathbf q}) \nonumber \\
%%%%%%%%%%%%%%%%%%%%%%  line-3
&&  + S \sum_{\mathbf q} \sum_{\gamma=x,y,z} J_{2,\gamma}\frac{1-\tau^{(2)}_{\gamma}}{2} e^{i{\mathbf q}\boldsymbol\delta^{(2)}_{\gamma}} (b_{\mathbf q}^{\dagger}b_{-\mathbf q}^{\dagger}) + \mathrm{h.c.}
\end{eqnarray}

Summing the contributions from both sublattices, the total second-neighbor Hamiltonian becomes
\begin{eqnarray}
\mathcal{H}_{2NN}
%%%%%%%%%%%%%%%%%%%%%%  line-1
&=& 2N_sS^2\sum_{\gamma=x,y,z} J_{2,\gamma}\tau^{(2)}_{\gamma}   \nonumber \\
%%%%%%%%%%%%%%%%%%%%%%  line-2
&&  + S \sum_{\mathbf q} \left(\sum_{\gamma=x,y,z} J_{2,\gamma}\frac{1+\tau^{(2)}_{\gamma}}{2} (e^{i{\mathbf q}\boldsymbol\delta^{(2)}_{\gamma}} + e^{-i{\mathbf q}\boldsymbol\delta^{(2)}_{\gamma}}) -2 \sum_{\gamma=x,y,z} J_{2,\gamma}\tau^{(2)}_{\gamma}\right) (a_{\mathbf q}^{\dagger}a_{\mathbf q} + b_{\mathbf q}^{\dagger}b_{\mathbf q}) \nonumber \\
%%%%%%%%%%%%%%%%%%%%%%  line-3
&&  + S \sum_{\mathbf q} \sum_{\gamma=x,y,z} J_{2,\gamma}\frac{1-\tau^{(2)}_{\gamma}}{2} e^{i{\mathbf q}\boldsymbol\delta^{(2)}_{\gamma}} (a_{\mathbf q}^{\dagger}a_{-\mathbf q}^{\dagger} + b_{\mathbf q}^{\dagger}b_{-\mathbf q}^{\dagger}) + \mathrm{h.c.}
\end{eqnarray}

\subsubsection{The third nearest-neighbor interaction $J_3$}

For the third-neighbor interactions, the Hamiltonian takes a similar structure. We find that
\begin{eqnarray}
\mathcal{H}_{3NN}
%%%%%%%%%%%%%%%%%%%%%%  line-1
&=& N_sS^2\sum_{\gamma} J_{3,\gamma}\tau^{(3)}_{\gamma} - S \sum_{\mathbf q} \sum_{\gamma=x,y,z} J_{3,\gamma}\tau^{(3)}_{\gamma} (a_{\mathbf q}^{\dagger}a_{\mathbf q} + b_{\mathbf q}^{\dagger}b_{\mathbf q})  \nonumber \\
%%%%%%%%%%%%%%%%%%%%%%  line-3
&&  + S \sum_{\mathbf q} \left(\sum_{\gamma} J_{3,\gamma}\frac{1+\tau^{(3)}_{\gamma}}{2} e^{i{\mathbf q}\boldsymbol\delta^{(3)}_{\gamma}}\right) a_{\mathbf q}^{\dagger}b_{\mathbf q} + \mathrm{h.c.} \nonumber \\
%%%%%%%%%%%%%%%%%%%%%%  line-4
&&  + S \sum_{\mathbf q} \left(\sum_{\gamma} J_{3,\gamma}\frac{1-\tau^{(3)}_{\gamma}}{2} e^{i{\mathbf q}\boldsymbol\delta^{(3)}_{\gamma}}\right) a_{\mathbf q}^{\dagger}b_{-\mathbf q}^{\dagger} + \mathrm{h.c.}
\end{eqnarray}

\subsubsection{The on-site single-ion anisotropy $A$}

Let us now concentrate on the single-ion anisotropy. Assuming that the spins are located in the $xz$ plane, the spin components along the crystallographic axes are related to the rotated frame via the angle $\theta$ between the classical spin direction and the $z$-axis. The $z$-component of the spin transforms as
\begin{eqnarray}
S_i^z = \cos \theta \, \tilde{S}_i^{z} - \sin \theta \, \tilde{S}_i^{x} = \tau (S - a_i^\dagger a_i) \cos\theta - \sqrt{\frac{S}{2}} (a_i + a_i^\dagger) \sin\theta
\end{eqnarray}
Similarly, the $x$-component is given by
\begin{eqnarray}
S_i^x = \sin \theta \, \tilde{S}_i^{z} + \cos \theta \, \tilde{S}_i^{x} = \tau (S - a_i^\dagger a_i) \sin\theta + \sqrt{\frac{S}{2}} (a_i + a_i^\dagger) \cos\theta
\end{eqnarray}

To obtain the quadratic spin-wave Hamiltonian, we square these expressions and keep terms up to second order in bosonic operators. For $(S_i^z)^2$, the expansion yields
\begin{eqnarray}
(S_i^z)^2
&\approx& (S - a_i^\dagger a_i)^2 \cos^2\theta + \frac{S}{2} (a_i + a_i^\dagger)^2 \sin^2\theta    \nonumber \\
&\approx& (S^2 - 2S a_i^\dagger a_i) \cos^2\theta + \frac{S}{2} [\underbrace{a_ia_i^\dagger}_{a_i^\dagger a_i+1} + a_i^\dagger a_i + a_i^2 + (a_i^\dagger)^2] \sin^2\theta    \nonumber \\
&\approx& (S \sin^2\theta - 2S \cos^2\theta) a_i^\dagger a_i + \frac{S}{2} (a_i^2 + (a_i^\dagger)^2) \sin^2\theta + \left(\frac{S}{2} \sin^2\theta + S^2 \cos^2\theta\right)  \nonumber \\
&\approx& S(1 - 3\cos^2\theta) a_i^\dagger a_i + \frac{S}{2} (a_i^2 + (a_i^\dagger)^2) \sin^2\theta + \left(\frac{S}{2} \sin^2\theta + S^2 \cos^2\theta\right)
\end{eqnarray}
Analogously, for $(S_i^x)^2$ we obtain
\begin{eqnarray}
(S_i^x)^2
&\approx& (S - a_i^\dagger a_i)^2 \sin^2\theta + \frac{S}{2} (a_i + a_i^\dagger)^2 \cos^2\theta    \nonumber \\
&\approx& (S^2 - 2S a_i^\dagger a_i) \sin^2\theta + \frac{S}{2} [\underbrace{a_ia_i^\dagger}_{a_i^\dagger a_i+1} + a_i^\dagger a_i + a_i^2 + (a_i^\dagger)^2] \cos^2\theta    \nonumber \\
&\approx& (S \cos^2\theta - 2S \sin^2\theta) a_i^\dagger a_i + \frac{S}{2} (a_i^2 + (a_i^\dagger)^2) \cos^2\theta + \left(\frac{S}{2} \cos^2\theta + S^2 \sin^2\theta\right)  \nonumber \\
&\approx& S(1 - 3\sin^2\theta) a_i^\dagger a_i + \frac{S}{2} (a_i^2 + (a_i^\dagger)^2) \cos^2\theta + \left(\frac{S}{2} \cos^2\theta + S^2 \sin^2\theta\right)
\end{eqnarray}

Finally, we combine the contributions from both $x$ and $z$ anisotropy terms. The single-ion anisotropy Hamiltonian $\mathcal{H}_{SIA} = \sum_i [A_x(S_i^x)^2 + A_z(S_i^z)^2]$ involves summation over all lattice sites. Since the two sublattices $\mathcal{A}$ and $\mathcal{B}$ are equivalent under the spin-wave transformation, we can sum over both and obtain the total contribution in momentum space

\begin{eqnarray}
\mathcal{H}_{SIA} &=& \sum_i [A_x(S_i^x)^2 + A_z(S_i^z)^2] = \sum_{i \in \mathcal{A}} + \sum_{i \in \mathcal{B}}  [A_x(S_i^x)^2 + A_z(S_i^z)^2]  \nonumber \\
%-------------------------------------------------------------------------------------
&\approx& 2N_s \cdot \left[A_z \left(\frac{S}{2} \sin^2\theta + S^2 \cos^2\theta\right) + A_x \left(\frac{S}{2} \cos^2\theta + S^2 \sin^2\theta\right) \right] \nonumber \\
&& + S \left[A_z (1 - 3\cos^2\theta) + A_x (1 - 3\sin^2\theta)\right] \sum_{{\mathbf q}} (a_{\mathbf q}^\dagger a_{\mathbf q} + b_{\mathbf q}^\dagger b_{\mathbf q} ) \nonumber \\
&& + \frac{S}{2} (A_z \sin^2\theta + A_x \cos^2\theta) \sum_{{\mathbf q}} \big(a_{\mathbf q}^\dagger a_{-\mathbf q}^\dagger + b_{\mathbf q}^\dagger b_{-\mathbf q}^\dagger\big) + \textrm{h.c.}
%-------------------------------------------------------------------------------------
\end{eqnarray}

\subsubsection{The full Hamiltonian}

Collecting all contributions from nearest, second-nearest, and third-nearest neighbor interactions, the total Hamiltonian can be expressed as
\begin{eqnarray}\label{HtotAACC}
\mathcal{H}
%%%%%%%%%%%%%%%%%%%%%%  line-0
&=& N_sS^2\left(\sum_{\gamma} J_{1,\gamma}\tau_{\gamma} + 2\sum_{\gamma=x,y,z} J_{2,\gamma}\tau^{(2)}_{\gamma} + \sum_{\gamma} J_{3,\gamma}\tau^{(3)}_{\gamma}\right) + N_sS^4\sum_{\gamma} B_{\gamma}   \nonumber \\
&& + 2N_s \cdot \left[A_z \left(\frac{S}{2} \sin^2\theta + S^2 \cos^2\theta\right) + A_x \left(\frac{S}{2} \cos^2\theta + S^2 \sin^2\theta\right) \right] \nonumber \\
%%%%%%%%%%%%%%%%%%%%%%  line-1
&&+ S \sum_{\mathbf q} \Bigg[ \left[A_z (1 - 3\cos^2\theta) + A_x (1 - 3\sin^2\theta)\right] - \sum_{\gamma=x,y,z} \left(J_{1,\gamma}\tau_{\gamma} + 2 J_{2,\gamma}\tau^{(2)}_{\gamma} + J_{3,\gamma}\tau^{(3)}_{\gamma} + 2S^2 B_{\gamma}\right) \nonumber \\
&&\qquad\qquad + \sum_{\gamma=x,y,z} J_{2,\gamma}\frac{1+\tau^{(2)}_{\gamma}}{2} \big(e^{i{\mathbf q}\boldsymbol\delta^{(2)}_{\gamma}} + e^{-i{\mathbf q}\boldsymbol\delta^{(2)}_{\gamma}}\big) \Bigg] \big( a_{\mathbf q}^{\dagger}a_{\mathbf q} + b_{\mathbf q}^{\dagger}b_{\mathbf q} \big)   \nonumber \\
%%%%%%%%%%%%%%%%%%%%%%  line-2
&&+ S \sum_{\mathbf q}
    \left(\sum_{\gamma=x,y,z} J^{+}_{1,\gamma} \frac{1+\tau_{\gamma}}{2} e^{i{\mathbf q}\boldsymbol\delta_{\gamma}} + \sum_{\gamma=x,y,z} J_{3,\gamma}\frac{1+\tau^{(3)}_{\gamma}}{2} e^{i{\mathbf q}\boldsymbol\delta^{(3)}_{\gamma}}\right)
    a_{\mathbf q}^{\dagger}b_{\mathbf q} + \textrm{h.c.}  \nonumber \\
%%%%%%%%%%%%%%%%%%%%%%  line-1
&&+ S \sum_{\mathbf q} \bigg[\sum_{\gamma=x,y,z} J_{2,\gamma}\frac{1-\tau^{(2)}_{\gamma}}{2} e^{i{\mathbf q}\boldsymbol\delta^{(2)}_{\gamma}} + \frac{A_z \sin^2\theta + A_x \cos^2\theta}{2}\bigg] (a_{\mathbf q}^{\dagger}a_{-\mathbf q}^{\dagger} + b_{\mathbf q}^{\dagger}b_{-\mathbf q}^{\dagger}) + \mathrm{h.c.} \nonumber \\
%%%%%%%%%%%%%%%%%%%%%%  line-2
&&+ S \sum_{\mathbf q} \left(\sum_{\gamma=x,y,z} J^{-}_{1,\gamma} \frac{1-\tau_{\gamma}}{2} e^{i{\mathbf q}\boldsymbol\delta_{\gamma}} + \sum_{\gamma=x,y,z} J_{3,\gamma}\frac{1-\tau^{(3)}_{\gamma}}{2} e^{i{\mathbf q}\boldsymbol\delta^{(3)}_{\gamma}}\right)
    a_{\mathbf q}^{\dagger}b_{-\mathbf q}^{\dagger} + \textrm{h.c.}  \nonumber \\
%%%%%%%%%%%%%%%%%%%%%%  line-777777777777777777777777777777777777
&=& N_s \cdot S(S+1) \left(\sum_{\gamma} J_{1,\gamma}\tau_{\gamma} + 2\sum_{\gamma} J_{2,\gamma}\tau^{(2)}_{\gamma} + \sum_{\gamma} J_{3,\gamma}\tau^{(3)}_{\gamma}\right) + N_s \cdot S^3(S+2)\sum_{\gamma} B_{\gamma}   \nonumber \\
& &  + 2N_s \cdot S(S+1) (A_z\cos^2\theta + A_x\sin^2\theta) - N_s \cdot S \sum_{\gamma=x,y,z} J_{2,\gamma}\frac{1+\tau^{(2)}_{\gamma}}{2} (e^{i{\mathbf q} \boldsymbol\delta^{(2)}_{\gamma}}+e^{-i{\mathbf q} \boldsymbol\delta^{(2)}_{\gamma}})   \nonumber \\
	&&+ \frac{S}{2} \sum_{\mathbf q}
    \hat{\bf x}_{\bf q}^\dagger
    \hat{\bf H}_{\bf q}
    \hat{\bf x}_{\bf q}^{\phantom{\dagger}}.
\end{eqnarray}
The final Hamiltonian is expressed in a compact matrix form. This representation is ideal for subsequent diagonalization via a Bogoliubov transformation, which yields the magnon excitation spectrum. The matrix $\hat{\bf H}_{\bf q}$ encodes all the interactions: the diagonal block $\hat{\Lambda}_{\bf q}$ contains the normal (number-conserving) terms, while the off-diagonal block $\hat{\Delta}_{\bf q}$ contains the anomalous (pairing) terms.
Here, $\hat{\bf x}^\dag_{\bf q}=\left( a^\dag_{\bf q}, b^\dag_{\bf k}, a^{\phantom \dag}_{\bf -q},b^{\phantom \dag}_{\bf -q}\right)$
is a vector of length $2 n_s$ and $\hat{\bf H}_{\bf k}$ is a $2 n_s\times 2 n_s$ matrix
\begin{eqnarray}\label{HtotAACC}
\hat{\bf H}_{\bf q} =
\left(\begin{array}{@{}cc@{}}
	\hat{\Lambda}_{\bf q}              &    \hat{\Delta}_{\bf q}       \\
    \hat{\Delta}_{\bf q}^{\dagger}     &    \hat{\Lambda}_{-\bf q}^T
\end{array}\right)
\end{eqnarray}
and
\begin{eqnarray}\label{HtotAACC}
\hat{\Lambda}_{\bf q} =
\left(\begin{array}{cccc}
	\mathbb{A}_{\mathbf q}                  & \mathbb{B}_{\mathbf q}      \\
    \mathbb{B}^{*}_{\mathbf q}  & \mathbb{A}_{\mathbf q}             \\
\end{array}\right),\quad
\hat{\Delta}_{\bf q} =
\left(\begin{array}{cccc}
	\mathbb{C}_{\mathbf q}       & \mathbb{D}_{\mathbf q}     \\
    \mathbb{D}_{-\mathbf q}      & \mathbb{C}_{\mathbf q}    \\
\end{array}\right)
\end{eqnarray}
where
\begin{align}
\left\{
  \begin{array}{lll}
    \mathbb{A}_{\mathbf q} &=& A_z (1 - 3\cos^2\theta) + A_x (1 - 3\sin^2\theta) - \sum_{\gamma=x,y,z} \left(J_{1,\gamma}\tau_{\gamma} + 2 J_{2,\gamma}\tau^{(2)}_{\gamma} + J_{3,\gamma}\tau^{(3)}_{\gamma} + 2S^2 B_{\gamma}\right) \nonumber \\
    && + \sum_{\gamma=x,y,z} J_{2,\gamma}\frac{1+\tau^{(2)}_{\gamma}}{2} (e^{i{\mathbf q} \boldsymbol\delta^{(2)}_{\gamma}}+e^{-i{\mathbf q} \boldsymbol\delta^{(2)}_{\gamma}}) \\
    \mathbb{B}_{\mathbf q} &=& \left(\sum_{\gamma=x,y,z} J^{+}_{1,\gamma} \frac{1+\tau_{\gamma}}{2} e^{i{\mathbf q} \boldsymbol\delta_{\gamma}} + \sum_{\gamma=x,y,z} J_{3,\gamma}\frac{1+\tau^{(3)}_{\gamma}}{2} e^{i{\mathbf q} \boldsymbol\delta^{(3)}_{\gamma}}\right)  \\
    \mathbb{C}_{\mathbf q} &=& \sum_{\gamma=x,y,z} J_{2,\gamma}\frac{1-\tau^{(2)}_{\gamma}}{2} e^{i{\mathbf q} \boldsymbol\delta^{(2)}_{\gamma}} + \frac{1}{2} (A_z \sin^2\theta + A_x \cos^2\theta)   \\
    \mathbb{D}_{\mathbf q} &=& \sum_{\gamma=x,y,z} J^{-}_{1,\gamma} \frac{1-\tau_{\gamma}}{2} e^{i{\mathbf q}\boldsymbol\delta_{\gamma}} + \sum_{\gamma=x,y,z} J_{3,\gamma}\frac{1-\tau^{(3)}_{\gamma}}{2} e^{i{\mathbf q} \boldsymbol\delta^{(3)}_{\gamma}}
  \end{array}
\right..
\end{align}
For the AFM phase, $(\tau_x, \tau_y, \tau_z) = (-1, -1, -1)$, $(\tau_x^{(2)}, \tau_y^{(2)}, \tau_z^{(2)}) = (+1, +1, +1)$, and $(\tau_x^{(3)}, \tau_y^{(3)}, \tau_z^{(3)}) = (-1, -1, -1)$. By using of these factors, we have
\begin{align}
\left\{
  \begin{array}{l}
    \mathbb{A}_{\mathbf q} = A_z (1 - 3\cos^2\theta) + A_x (1 - 3\sin^2\theta) - \left(-3J_{1} + 6J_{2} - 3J_{3} + 6S^2 B\right) + J_{2} \sum_{\gamma=x,y,z} (e^{i{\mathbf q} \boldsymbol\delta^{(2)}_{\gamma}}+e^{-i{\mathbf q} \boldsymbol\delta^{(2)}_{\gamma}}) \\
    \mathbb{B}_{\mathbf q} = 0  \\
    \mathbb{C}_{\mathbf q} = \frac{1}{2} (A_z \sin^2\theta + A_x \cos^2\theta)   \\
    \mathbb{D}_{\mathbf q} = J^{-}_{1}\sum_{\gamma=x,y,z} e^{i{\mathbf q}\boldsymbol\delta_{\gamma}} + J_{3}\sum_{\gamma=x,y,z} e^{i{\mathbf q} \boldsymbol\delta^{(3)}_{\gamma}}
  \end{array}
\right..
\end{align}

\subsection{Four-sublattice phases: zigzag phase as an example}

The derivation of the BdG Hamiltonian for the zigzag phase proceeds analogously to the AFM case, though it is considerably more involved. Without delving into the details, we present the spin-wave Hamiltonian as
\begin{eqnarray}\label{HtotAACC}
\mathcal{H}
%%%%%%%%%%%%%%%%%%%%%%  line-0
&=& 2N_sS^2 \left(\sum_{\gamma} J_{1,\gamma}\tau_{\gamma} + 2\sum_{\gamma} J_{2,\gamma}\tau^{(2)}_{\gamma} + \sum_{\gamma} J_{3,\gamma}\tau^{(3)}_{\gamma}\right) +  2N_sS^4\sum_{\gamma} B_{\gamma}   \nonumber \\
&& + 4N_s \cdot \left[A_z \left(\frac{S}{2} \sin^2\theta + S^2 \cos^2\theta\right) + A_x \left(\frac{S}{2} \cos^2\theta + S^2 \sin^2\theta\right) \right]  \nonumber \\
%%%%%%%%%%%%%%%%%%%%%%  line-1
&&+ S \sum_{\mathbf q} \Bigg[\left[A_z (1 - 3\cos^2\theta) + A_x (1 - 3\sin^2\theta)\right] - \sum_{\gamma=x,y,z} \left(J_{1,\gamma}\tau_{\gamma} + 2 J_{2,\gamma}\tau^{(2)}_{\gamma} + J_{3,\gamma}\tau^{(3)}_{\gamma} + 2S^2 B_{\gamma}\right) \nonumber \\
&&\qquad\qquad + \sum_{\gamma=z} J_{2,\gamma}\frac{1+\tau^{(2)}_{\gamma}}{2} \big(e^{i{\mathbf q}\boldsymbol\delta^{(2)}_{\gamma}} + e^{-i{\mathbf q}\boldsymbol\delta^{(2)}_{\gamma}}\big) \Bigg] \big( a_{\mathbf q}^{\dagger}a_{\mathbf q} + b_{\mathbf q}^{\dagger}b_{\mathbf q}
    + c_{\mathbf q}^{\dagger}c_{\mathbf q} + d_{\mathbf q}^{\dagger}d_{\mathbf q} \big)   \nonumber \\
%%%%%%%%%%%%%%%%%%%%%%  line-2
&&+ S \sum_{\mathbf q}
    \left(\sum_{\gamma=z} J^{+}_{1,\gamma} \frac{1+\tau_{\gamma}}{2} e^{i{\mathbf q}\boldsymbol\delta_{\gamma}} + \sum_{\gamma=x,y,z} J_{3,\gamma}\frac{1+\tau^{(3)}_{\gamma}}{2} e^{i{\mathbf q}\boldsymbol\delta^{(3)}_{\gamma}}\right)
    (a_{\mathbf q}^{\dagger}b_{\mathbf q} + c_{\mathbf q}^{\dagger}d_{\mathbf q}) + \textrm{h.c.}  \nonumber \\
%%%%%%%%%%%%%%%%%%%%%%  line-3
&&+ S \sum_{\mathbf q} \sum_{\gamma=x,y} J_{2,\gamma}\frac{1+\tau^{(2)}_{\gamma}}{2} \big(e^{i{\mathbf q}\boldsymbol\delta^{(2)}_{\gamma}} + e^{-i{\mathbf q}\boldsymbol\delta^{(2)}_{\gamma}}\big) (a_{\mathbf q}^{\dagger}c_{\mathbf q} + b_{\mathbf q}^{\dagger}d_{\mathbf q}) + \mathrm{h.c.} \nonumber \\
%%%%%%%%%%%%%%%%%%%%%%  line-4
&&+ S \sum_{\mathbf q} \sum_{\gamma=x,y} J^{+}_{1,\gamma} \frac{1+\tau_{\gamma}}{2} e^{i{\mathbf q}\boldsymbol\delta_{\gamma}}
        (a_{\mathbf q}^{\dagger}d_{\mathbf q} + c_{\mathbf q}^{\dagger}b_{\mathbf q}) + \textrm{h.c.}  \nonumber \\
%%%%%%%%%%%%%%%%%%%%%%  line-1
&&+ S \sum_{\mathbf q} \bigg[\sum_{\gamma=z} J_{2,\gamma}\frac{1-\tau^{(2)}_{\gamma}}{2} e^{i{\mathbf q}\boldsymbol\delta^{(2)}_{\gamma}} + \frac{A_z \sin^2\theta + A_x \cos^2\theta}{2}\bigg] (a_{\mathbf q}^{\dagger}a_{-\mathbf q}^{\dagger} + b_{\mathbf q}^{\dagger}b_{-\mathbf q}^{\dagger} + c_{\mathbf q}^{\dagger}c_{-\mathbf q}^{\dagger} + d_{\mathbf q}^{\dagger}d_{-\mathbf q}^{\dagger}) + \mathrm{h.c.} \nonumber \\
%%%%%%%%%%%%%%%%%%%%%%  line-2
&&+ S \sum_{\mathbf q} \left(\sum_{\gamma=z} J^{-}_{1,\gamma} \frac{1-\tau_{\gamma}}{2} e^{i{\mathbf q}\boldsymbol\delta_{\gamma}} + \sum_{\gamma=x,y,z} J_{3,\gamma}\frac{1-\tau^{(3)}_{\gamma}}{2} e^{i{\mathbf q}\boldsymbol\delta^{(3)}_{\gamma}}\right)
    (a_{\mathbf q}^{\dagger}b_{-\mathbf q}^{\dagger} + c_{\mathbf q}^{\dagger}d_{-\mathbf q}^{\dagger}) + \textrm{h.c.}  \nonumber \\
%%%%%%%%%%%%%%%%%%%%%%  line-3
&&+ S \sum_{\mathbf q} \sum_{\gamma=x,y} J_{2,\gamma}\frac{1-\tau^{(2)}_{\gamma}}{2} \big(e^{i{\mathbf q}\boldsymbol\delta^{(2)}_{\gamma}} + e^{-i{\mathbf q}\boldsymbol\delta^{(2)}_{\gamma}}\big) (a_{\mathbf q}^{\dagger}c_{-\mathbf q}^{\dagger} + b_{\mathbf q}^{\dagger}d_{-\mathbf q}^{\dagger}) + \mathrm{h.c.} \nonumber \\
%%%%%%%%%%%%%%%%%%%%%%  line-4
&&+ S \sum_{\mathbf q} \sum_{\gamma=x,y} J^{-}_{1,\gamma} \frac{1-\tau_{\gamma}}{2} e^{i{\mathbf q}\boldsymbol\delta_{\gamma}}
        (a_{\mathbf q}^{\dagger}d_{-\mathbf q}^{\dagger} + c_{\mathbf q}^{\dagger}b_{-\mathbf q}^{\dagger}) + \textrm{h.c.}  \nonumber \\
%%%%%%%%%%%%%%%%%%%%%%  line-777777777777777777777777777777777777
&=& 2N_s \cdot S(S+1) \left(\sum_{\gamma} J_{1,\gamma}\tau_{\gamma} + 2\sum_{\gamma} J_{2,\gamma}\tau^{(2)}_{\gamma} + \sum_{\gamma} J_{3,\gamma}\tau^{(3)}_{\gamma}\right) + 2N_s \cdot S^3(S+2)\sum_{\gamma} B_{\gamma}   \nonumber \\
& &  + 4N_s \cdot S(S+1) (A_z\cos^2\theta + A_x\sin^2\theta) - 2N_s \cdot S \sum_{\gamma=z} J_{2,\gamma}\frac{1+\tau^{(2)}_{\gamma}}{2} (e^{i{\mathbf q} \boldsymbol\delta^{(2)}_{\gamma}}+e^{-i{\mathbf q} \boldsymbol\delta^{(2)}_{\gamma}})   \nonumber \\
	&&+ \frac{S}{2} \sum_{\mathbf q}
    \hat{\bf x}_{\bf q}^\dagger
    \hat{\bf H}_{\bf q}
    \hat{\bf x}_{\bf q}^{\phantom{\dagger}}.
\end{eqnarray}
where $\hat{\bf x}^\dag_{\bf q}=\left( a^\dag_{\bf q}, b^\dag_{\bf k}, c^\dag_{\bf q}, d^\dag_{\bf k},
a^{\phantom \dag}_{\bf -q},b^{\phantom \dag}_{\bf -q}, c^{\phantom \dag}_{\bf -q}, d^{\phantom \dag}_{\bf -q}\right)$
is a vector of length $2 n_s$ and $\hat{\bf H}_{\bf k}$ is a $2 n_s\times 2 n_s$ matrix
\begin{eqnarray}\label{HtotAACC}
\hat{\bf H}_{\bf q} =
\left(\begin{array}{@{}cc@{}}
	\hat{\Lambda}_{\bf q}              &    \hat{\Delta}_{\bf q}       \\
    \hat{\Delta}_{\bf q}^{\dagger}     &    \hat{\Lambda}_{-\bf q}^T
\end{array}\right)
\end{eqnarray}
and
\begin{eqnarray}\label{HtotAACC}
\hat{\Lambda}_{\bf q} =
\left(\begin{array}{cccc}
	\mathbb{A}_{\mathbf q}                  & \mathbb{B}_{\mathbf q}      & \mathbb{C}_{\mathbf q}         & \mathbb{D}_{\mathbf q}     \\
    \mathbb{B}^{*}_{\mathbf q}  & \mathbb{A}_{\mathbf q}                  & \mathbb{D}^{*}_{\mathbf q}     & \mathbb{C}_{\mathbf q}     \\
    \mathbb{C}^{*}_{\mathbf q}  & \mathbb{D}_{\mathbf q}      & \mathbb{A}_{\mathbf q}                     & \mathbb{B}_{\mathbf q}     \\
    \mathbb{D}^{*}_{\mathbf q}  & \mathbb{C}^{*}_{\mathbf q}  & \mathbb{B}^{*}_{\mathbf q}     & \mathbb{A}_{\mathbf q}                 \\
\end{array}\right),\quad
\hat{\Delta}_{\bf q} =
\left(\begin{array}{cccc}
	\mathbb{E}_{\mathbf q}       & \mathbb{F}_{\mathbf q}     &  \mathbb{G}_{\mathbf q}        & \mathbb{H}_{\mathbf q}     \\
    \mathbb{F}_{-\mathbf q}      & \mathbb{E}_{\mathbf q}     &  \mathbb{H}_{-\mathbf q}       & \mathbb{G}_{\mathbf q}     \\
    \mathbb{G}_{-\mathbf q}      & \mathbb{H}_{\mathbf q}     &  \mathbb{E}_{\mathbf q}        & \mathbb{F}_{\mathbf q}     \\
    \mathbb{H}_{-\mathbf q}      & \mathbb{G}_{-\mathbf q}    &  \mathbb{F}_{-\mathbf q}       & \mathbb{E}_{\mathbf q}     \\
\end{array}\right)
\end{eqnarray}
where
\begin{align}
\left\{
  \begin{array}{lll}
    \mathbb{A}_{\mathbf q} &=& A_z (1 - 3\cos^2\theta) + A_x (1 - 3\sin^2\theta) - \sum_{\gamma=x,y,z} \left(J_{1,\gamma}\tau_{\gamma} + 2 J_{2,\gamma}\tau^{(2)}_{\gamma} + J_{3,\gamma}\tau^{(3)}_{\gamma} + 2S^2 B_{\gamma}\right) \nonumber \\
    && + \sum_{\gamma=z} J_{2,\gamma}\frac{1+\tau^{(2)}_{\gamma}}{2} (e^{i{\mathbf q} \boldsymbol\delta^{(2)}_{\gamma}}+e^{-i{\mathbf q} \boldsymbol\delta^{(2)}_{\gamma}}) \\
    \mathbb{B}_{\mathbf q} &=& \left(\sum_{\gamma=z} J^{+}_{1,\gamma} \frac{1+\tau_{\gamma}}{2} e^{i{\mathbf q} \boldsymbol\delta_{\gamma}} + \sum_{\gamma=x,y,z} J_{3,\gamma}\frac{1+\tau^{(3)}_{\gamma}}{2} e^{i{\mathbf q} \boldsymbol\delta^{(3)}_{\gamma}}\right)  \\
    \mathbb{C}_{\mathbf q} &=& \sum_{\gamma=x,y} J_{2,\gamma}\frac{1+\tau^{(2)}_{\gamma}}{2} \big(e^{i{\mathbf q} \boldsymbol\delta^{(2)}_{\gamma}} + e^{-i{\mathbf q} \boldsymbol\delta^{(2)}_{\gamma}}\big)  \\
    \mathbb{D}_{\mathbf q} &=& \sum_{\gamma=x,y} J^{+}_{1,\gamma} \frac{1+\tau_{\gamma}}{2} e^{i{\mathbf q} \boldsymbol\delta_{\gamma}} \\
    \mathbb{E}_{\mathbf q} &=& \sum_{\gamma=z} J_{2,\gamma}\frac{1-\tau^{(2)}_{\gamma}}{2} e^{i{\mathbf q} \boldsymbol\delta^{(2)}_{\gamma}} + \frac{1}{2} (A_z \sin^2\theta + A_x \cos^2\theta)   \\
    \mathbb{F}_{\mathbf q} &=& \sum_{\gamma=z} J^{-}_{1,\gamma} \frac{1-\tau_{\gamma}}{2} e^{i{\mathbf q}\boldsymbol\delta_{\gamma}} + \sum_{\gamma=x,y,z} J_{3,\gamma}\frac{1-\tau^{(3)}_{\gamma}}{2} e^{i{\mathbf q} \boldsymbol\delta^{(3)}_{\gamma}} \\
    \mathbb{G}_{\mathbf q} &=& \sum_{\gamma=x,y} J_{2,\gamma}\frac{1-\tau^{(2)}_{\gamma}}{2} \big(e^{i{\mathbf q}\boldsymbol\delta^{(2)}_{\gamma}} + e^{-i{\mathbf q} \boldsymbol\delta^{(2)}_{\gamma}}\big)  \\
    \mathbb{H}_{\mathbf q} &=& \sum_{\gamma=x,y} J^{-}_{1,\gamma} \frac{1-\tau_{\gamma}}{2} e^{i{\mathbf q}\boldsymbol\delta_{\gamma}}
  \end{array}
\right..
\end{align}
For the zigzag phase, $(\tau_x, \tau_y, \tau_z) = (+1, +1, -1)$, $(\tau_x^{(2)}, \tau_y^{(2)}, \tau_z^{(2)}) = (-1, -1, +1)$, and $(\tau_x^{(3)}, \tau_y^{(3)}, \tau_z^{(3)}) = (-1, -1, -1)$. By using of these factors, we have
\begin{align}
\left\{
  \begin{array}{l}
    \mathbb{A}_{\mathbf q} = A_z (1 - 3\cos^2\theta) + A_x (1 - 3\sin^2\theta) - \left(J_{1} - 2J_{2} - 3J_{3} + 6S^2 B\right) + J_{2} (e^{i{\mathbf q} \boldsymbol\delta^{(2)}_{\gamma}}+e^{-i{\mathbf q} \boldsymbol\delta^{(2)}_{\gamma}}) \\
    \mathbb{B}_{\mathbf q} = 0  \\
    \mathbb{C}_{\mathbf q} = 0  \\
    \mathbb{D}_{\mathbf q} = J^{+}_{1,\gamma} \sum_{\gamma=x,y} e^{i{\mathbf q} \boldsymbol\delta_{\gamma}} \\
    \mathbb{E}_{\mathbf q} = \frac{1}{2} (A_z \sin^2\theta + A_x \cos^2\theta)   \\
    \mathbb{F}_{\mathbf q} = J^{-}_{1} e^{i{\mathbf q}\boldsymbol\delta_{3}} + J_{3}\sum_{\gamma=x,y,z} e^{i{\mathbf q} \boldsymbol\delta^{(3)}_{\gamma}} \\
    \mathbb{G}_{\mathbf q} = J_{2} \sum_{\gamma=x,y}  \big(e^{i{\mathbf q}\boldsymbol\delta^{(2)}_{\gamma}} + e^{-i{\mathbf q} \boldsymbol\delta^{(2)}_{\gamma}}\big)  \\
    \mathbb{H}_{\mathbf q} = 0
  \end{array}
\right..
\end{align}

%%%%%%%%%%%%%%%%%%%%%%%%%%%%%%%%%%%%%%%%%%%%%%%%%%%%%%%%%%%%%%%%%%%%%%%%%%%%%%%%%%%%%%%%%%%%%%%%%%%%%%%%%%%%%%%%%%%%%%%%%%%%%%%%%%%%%%%%%%%%%%%%
% \newpage

%
%%%%%%%%%%%%%%%%%%%%%%%%%%%%%%%%%%%%%%%%%%%%%%%%%%%%%%%%%%%%%%%%%%%%%%%%%%

% \end{CJK*}

\end{document}